# On Stability Region and Delay Performance of Linear-Memory Randomized Scheduling for Time-Varying Networks

Mahdi Lotfinezhad, Ben Liang, Elvino S. Sousa

*Abstract*—**Throughput optimal scheduling policies in general require the solution of a complex and often NP-hard optimization problem. Related literature has shown that in the context of time-varying channels, randomized scheduling policies can be employed to reduce the complexity of the optimization problem but at the expense of a memory requirement that is exponential in the number of data flows. In this paper, we consider a Linear-Memory Randomized Scheduling Policy (LM-RSP) that is based on a pick-and-compare principle in a time-varying network with $N$ one-hop data flows. For general ergodic channel processes, we study the performance of LM-RSP in terms of its stability region and average delay. Specifically, we show that LM-RSP can stabilize a fraction of the capacity region. Our analysis characterizes this fraction as well as the average delay as a function of channel variations and the efficiency of LM-RSP in choosing an appropriate schedule vector. Applying these results to a class of Markovian channels, we provide explicit results on the stability region and delay performance of LM-RSP.**

## I. INTRODUCTION

One key characteristic of the wireless communication medium is its random variations due to user mobility and unpredictable changes in the radio environment. This makes the enduring challenge of efficient resource scheduling extremely difficult in wireless networks, especially as the network size increases. In their seminal work [1], Tassiulas and Ephremides propose a *throughput optimal* scheduling policy, commonly referred to as the Generalized Maximum Weight Matching (GMWM) policy, that stabilizes the network for *any* input rate that is within the *network layer capacity region*. In this context, the network layer capacity region is defined as the closure of the set of all input rates that can be *stably* supported by the network using any possible scheduling policy [1][2][3].

The GMWM policy in each time-slot maximizes the sum of backlog-rate products given the channel states and queue-lengths, where this maximization can be considered as a GMWM problem, which can be NP-hard depending on the underlying interference model [4]. The complexity of the GMWM policy naturally has motivated many researchers to develop sub-optimal algorithms that approximate its solution. In particular, Tassiulas in a pioneering work [5] shows that simple randomized policies based on the *pick-and-compare* principle are sufficient to achieve throughput optimality. These policies in each time-slot use a *randomized* algorithm $A$ to select a *candidate* schedule vector that with non-zero probability $\delta$ can be the optimal solution to the GMWM problem. Once a schedule vector is picked, it is compared with the previous schedule in terms of the sum of backlog-rate products, and the one with the larger sum is selected for scheduling. This approach, however, assumes time-invariant channels.

More recently, in the context of time-varying channels, the authors of [6][7] have shown throughput optimality can be achieved if at the generic time-slot $t$, the randomized policy compares the picked schedule vector with the one used at the most recent occurrence of the same channel state at time-slot $t$. This proposal, therefore, requires a table with a size proportional to the number of channel states, and hence, its memory requirement is exponential in the number of data flows (links) $N$. In practice, however, mobile systems are computationally limited and have limited memory resource. Therefore, any attempt to practically implement such randomized policies should aim at reducing both the complexity of computation and the required memory storage.

In this paper, we are interested in addressing the following questions under the assumption that randomized policies are employed for scheduling:

- *How much sub-optimality in the network throughput is introduced by a reduced memory requirement, especially when the available memory storage can increase only linearly with the number of data flows?*
- *What is the delay scaling? Does the delay increase exponentially as the channel states become increasingly correlated and the number of data flows increase?*

Inspired by the above challenging questions, in this paper, we focus on a linear-memory randomized scheduling policy LM-RSP, which essentially follows the same pick-and-compare principle as the one used in the randomized policies in [5][6][7] but is generalized in the following respects. First, the update rule for the comparison of schedules in LM-RSP is generalized to be probabilistic. Second, LM-RSP uses a more general model for the randomized algorithm $A$, according to which, with a probability not less than a positive $\delta$, the algorithm $A$ returns a schedule vector that is within $\zeta$-*neighborhood* of the optimal solution. Considering different values for the pair $(\zeta, \delta)$ allows us to study algorithms with a wide range of complexity levels. Note that a value of $\delta$ less than one allows us to model algorithms with nondeterministic results, possibly those implemented in a distributed manner. In this paper, we limit our model to a network with $N$ one-hop data flows, e.g, downlink or uplink of a cellular or a mesh network.

Our main contribution in this paper is to analytically characterize the performance of LM-RSP in terms of its associated stability region and average delay in the context of time-varying channels. *First*, for general ergodic channel processes, we show that the stability region is a *scaled version* (fraction) of the network layer capacity region. Our analysis quantifies the scaling factor and demonstrates how it changes with *channel variations* and the *computational efficiency* of the randomized algorithm $A$. In addition, our analysis provides a general average delay-bound for the input rates strictly inside the studied stability region.

*Second*, to obtain more specific results, we consider an important class of Markovian channels where the state of each link is a two-state Markov chain. We assume that a link holds its state during one timeslot, but the state may change from one timeslot to another with (transition) probability $r$ and independent of the states of other links in the network. For this simple yet worst-case modeling class of channels, $r$ represents the *individual* link variation rate over one timeslot, and we show that for appropriate choice of parameters while the average delay is $\mathcal{O}(\frac{1}{r^2})$, as $r \to 0$, LM-RSP can stabilize a minimum fraction $\frac{\delta}{\delta+r}(1-\zeta)$ of the capacity region, e.g., when the interference is specified by the node exclusive model [8][4][9][10] or, more generally, by the $\kappa$-hop interference model [4], where no two links within $\kappa$ hops can successfully transmit at the same time. It is worth mentioning that this minimum fraction does not depend on the *total* channel variation rate, which approximately equals $Nr$ for small $r$'s, but, instead, depends on the individual link variation rate $r$. In addition, note that while the capacity region shrinks as the interference becomes more restrictive, e.g., when $\kappa$ increases in the $\kappa$-hop interference model, these results indicate that the minimum fraction remains fixed. Another important yet intuitive implication of these results is that if it is possible to increase $\delta$, at the expense of increasing the complexity of algorithm $A$, it is sufficient to make sure that $\delta$ has the same order as $r$ in order to make sure that LM-RSP stabilizes a fraction close to $(1-\zeta)$ of the capacity region.

Our results further verify that the average delay can be polynomially bounded as the number of data flows increases, e.g., when channels are Markovian, as described earlier. As far as we are aware, our results are the first to rigorously show that the delay does not need to increase exponentially with the number of data flows or channel correlation when randomized policies are used in the context of time-varying channels. Finally, note that in the limit of highly correlated channels, our results include the one in [5], which states that linear-complexity algorithms are sufficient to attain throughput optimality.

The rest of this paper is organized as follows. In Section II, we review the related work. In Section III, we provide details about the system model, and, in Section IV, we explain the operation of LM-RSP. In Sections V and VI, we state the main results of the paper on LM-RSP's stability region and delay performance, respectively. In Section VII, we discuss important observations. Finally, in Section VIII, we conclude the paper.

## II. RELATED WORK

Stable resource control for wireless networks first appeared in [1], where both the GMWM policy and the throughput capacity region are characterized. The work in [1], however, uses many simplifying assumptions for the channel and arrival processes. This result has been further extended by many researchers [11][2][12][6][3]. Recently, Neely and Modiano established the network layer capacity region for general ergodic channels and arrival processes [3]. In fact, the same stabilizing rule in [1] has been used as the scheduling component for rate control [13], energy optimal design [14], congestion control [15][16], and the study of utility-delay tradeoffs [17]. All of these papers generally assume that the scheduler has access to the solution of the complex GMWM problem. Another example of the throughput optimal control is the exponential rule proposed in [18].

One difficulty in implementing the GMWM policy is having access to updated queue information. In [6], it is shown that under general conditions, delayed or infrequent queue-length information does not affect the stability region. A similar result is shown to hold [19], when the queue-length-based scheduling at the base station is combined with congestion control at the end user, which can lead to weighted proportional fairness [15]. In this paper, therefore, we assume that the queue information is available and, instead, focus on the memory requirement and the complexity of the scheduling policies.

The main difficulty in implementing the GMWM policy is its complexity since this policy can be NP-hard depending on the assumed interference model [4]. This has motivated many researchers to develop sub-optimal constant-factor approximations to the GMWM policy. For instance, in [8], the impact of *imperfect* schedules is studied, where, as an example, a Maximal Matching (MM) scheduling algorithm is used to stabilize at least half of the capacity region. Due to its simplicity of implementation, MM scheduling has been widely investigated in the literature [20][4][21][10][22]. Despite the fact that these works address the issue of complexity, they are generally proposed for networks with time-invariant channels, or otherwise, do not exploit the channel correlation to improve the scheduling performance.

The use of randomized policies, based on the pick-and-compare principle, to reduce the complexity of throughput optimal scheduling first appeared in [5]. In a more recent work [9], the authors propose distributed schemes to implement a randomized policy similar to the one in [5] that can stabilize the entire capacity region. Both policies in [5] and [9], however, are proposed for time-invariant channels. In the context of time-varying channels, other proposals that are based on the policy in [5] include [6][7]. Although these proposals are throughput optimal, their memory requirement is exponential in the number of data flows, and thus, they may not be amenable to practical implementation in large networks.

In a different context, dynamic rate allocation has been proposed in [23]. This algorithm normalizes the feasible rates by appropriate weights and chooses the user with the largest normalized rate. Opportunistic scheduling is used in [24][25][26], where the long term fairness is achieved by assigning offsets to the users' utility functions [24], or by dynamically updating throughput weights [25][26]. MaxMin fair scheduling is studied in [27]. User level performance of channel-aware scheduling algorithms has been investigated in [28]. Asymptotic properties of the proportional fair sharing algorithms are studied in [29]. Our work in this paper does not consider the issue of fairness. Instead, we focus on throughput optimality and investigate the impact of channel variations on the performance of linear-memory low-complexity randomized scheduling policies.

## III. SYSTEM MODEL

We consider a wireless network with $N$ one-hop source-destination pairs, where each pair represents a data flow[1]. Associated with each data flow, we consider a separate queue, maintained at the source of the flow, which holds packets to be transmitted over a wireless link. Examples of this type of network include downlink or uplink of a cellular or a mesh network.

---

[1] Extension to multi-hop flows is possible using the methods in [1][3].

### A. Queueing

We assume the system is time-slotted, and channels hold their state during a time-slot but may change from one time-slot to another. Let $\mathbf{s}(t)$ be the matrix of all channel states from any given node $i$ to any other node $j$ in the network at time $t$. For instance, in the downlink of a cellular network, $\mathbf{s}(t)$ will reduce to the vector of user-base-station channel states, i.e., $\mathbf{s}(t) = (s_1(t), \ldots, s_N(t))$, where $s_i(t)$ is the state of the $i_{\text{th}}$ link at time $t$. Throughout the paper, we use bold face to denote vectors or matrices. Let $\mathcal{S}$ represent the set of all possible channel state matrices with finite cardinality $|\mathcal{S}|$. Let $D_i(t)$ denote the discrete rate over the $i_{\text{th}}$ link at time $t$, and $\mathbf{D}(t)$ be the corresponding vector of rates, i.e., $\mathbf{D}(t) = (D_1(t), \ldots, D_N(t))$. In addition, let $I_i(t)$ represent the amount of resource used by the $i_{\text{th}}$ link at time $t$, and $\mathbf{I}(t)$ be the corresponding vector, i.e., $\mathbf{I}(t) = (I_1(t), \cdots, I_N(t))$. The vector $\mathbf{I}(t)$ contains both scheduling and resource usage information, and hereafter, we refer to it simply as the *schedule vector*. Details for the selection of $\mathbf{I}(t)$ are provided in Section IV. Let $\mathcal{I}$ denote the set containing all possible schedule vectors, with finite cardinality $|\mathcal{I}|$.

Note that the exact specification of the schedule vector $\mathbf{I}(t)$ is system dependent. For instance, in CDMA systems, it may represent the vector of power levels associated with wireless links, or when the interference is characterized by the $\kappa$-hop interference model [4], the vector $\mathbf{I}(t)$ can be an activation vector representing a sub-graph in the network.

Since transmission rates are completely characterized given the channel states, the schedule vector, and the interference model, we have

$$\mathbf{D}(t) = \mathbf{D}(\mathbf{s}(t), \mathbf{I}(t)). \qquad (1)$$

We assume that transmission rates are bounded, i.e., for all $\mathbf{s} \in \mathcal{S}$ and $\mathbf{I} \in \mathcal{I}$,

$$D_l(\mathbf{s}, \mathbf{I}) < D_{max}, \; 1 \leq l \leq N, \qquad (2)$$

for some large $D_{max} > 0$.

Let $A_l(t)$ be the number of packets arriving in time-slot $t$ associated with the $l_{\text{th}}$ link (or data flow), and $\mathbf{A}(t)$ be the vector of arrivals, i.e., $\mathbf{A}(t) = (A_1(t), \cdots, A_N(t))$. We assume arrivals are i.i.d.[2] with mean vector

$$\mathbb{E}[\mathbf{A}(t)] = \mathbf{a} = (a_1, \ldots, a_N),$$

and with finite second moments:

$$\mathbb{E}[A_l^2(t)] < \tilde{A}_{max}^2, \; 1 \leq l \leq N, \qquad (3)$$

for a suitably large $\tilde{A}_{max}$. Assuming $\|\cdot\|$ represents the Euclidean norm of a vector, we define $\mathbb{E}[\|\mathbf{A}\|^2]$ as

$$\mathbb{E}[\|\mathbf{A}\|^2] = \mathbb{E}[\|\mathbf{A}(t)\|^2] = \sum_{l=1}^{N} \mathbb{E}[A_l^2(t)]. \qquad (4)$$

Since the arrival process is i.i.d., we see that $\mathbb{E}[\|\mathbf{A}\|^2]$ is well-defined and is independent of $t$. By Markov's inequality, we have the following fact.

**Fact 1.** *For any given positive $\epsilon$, there exists a sufficiently large $A_\epsilon$ such that for $A \geq A_\epsilon$, we have $p(\|\mathbf{A}(t)\| > A) < \epsilon$.*

Finally, let $\mathbf{X}(\mathbf{t}) = (X_1(t), \ldots, X_N(t))$ be the discrete vector of queue-lengths, where $X_l(t)$ is the queue-length associated with the $l_{\text{th}}$ link. Using the preceding definitions, we see that $\mathbf{X}(t)$ evolves according to the following equation

$$\mathbf{X}(t+1) = \mathbf{X}(t) + \mathbf{A}(t) - \mathbf{D}(t) + \mathbf{U}(t), \qquad (5)$$

where $\mathbf{U}(t)$ represents the wasted service vector with non-negative elements; the service is wasted when in a queue the number of packets waiting for transmission is less than the number that can be transmitted, i.e., when $X_l(t) < D_l(t)$ for some $l$, $1 \leq l \leq N$.

### B. Channel State Process

We assume the channel state process is stationary and ergodic. In particular, similar to [3], we assume for any given positive $\epsilon$, there exists a $K_{1,\epsilon}$ such that for $K \geq K_{1,\epsilon}$ regardless of the *system state* at time $t$ denoted by

$$\mathbf{Y}(t) = (\mathbf{X}(t), \mathbf{I}(t), \mathbf{s}(t)),$$

we have

$$\sum_{\mathbf{s} \in \mathcal{S}} \left| \pi(\mathbf{s}) - \mathbb{E}\left[ \frac{1}{K} \sum_{k=0}^{K-1} \mathbf{1}_{\mathbf{s}(t+k) = \mathbf{s}} \big| \mathbf{Y}(t) \right] \right| < \epsilon, \qquad (6)$$

where $\pi(\mathbf{s})$ is the *steady-state* probability of the channel at state $\mathbf{s}$, and $\mathbf{1}_e$ is the indicator function for the event $e$. This inequality simply states that the expected value of the average number of visits to a given channel state converges to its steady-state probability, and the sum of the absolute value of the differences, over all possible states, vanishes for sufficiently long time-intervals. The above further implies that there exists a $K_{2,\epsilon}^{(\gamma)}$ such that for $K \geq K_{2,\epsilon}^{(\gamma)}$, we have

$$\sum_{\mathbf{s} \in \mathcal{S}} \left| \pi(\mathbf{s}) - \mathbb{E}\left[ \frac{1}{\sum_{k=0}^{K-1} \gamma_{K-k}} \sum_{k=0}^{K-1} \gamma_{K-k} \mathbf{1}_{\mathbf{s}(t+k) = \mathbf{s}} \big| \mathbf{Y}(t) \right] \right| < \epsilon,$$

where $\{\gamma_i\}_{i=1}^{\infty}$ is an increasing sequence of positive real numbers with the property $\lim_{i \to \infty} \gamma_i = \gamma_\infty < \infty$. Note that $K_{2,\epsilon}^{(\gamma)}$ in general depends on the sequence $\{\gamma_i\}_{i=0}^{\infty}$. Of particular interest is the case where $\gamma_i = \sum_{j=1}^{i} (1-\delta)^j$, which defines the sequence $\{\gamma_i\}_{i=1}^{\infty}$ for the rest of the paper. Examples of processes that satisfy the above inequalities include but are not limited to Markov chains.

## IV. SCHEDULING POLICY

In this section, we elaborate on the statistical structure of the algorithms that provide sub-optimal solutions to the GMWM problem, and describe the operation of LM-RSP that uses these algorithms. We start by providing a brief overview of the network layer capacity region and the precise definition of the GMWM problem.

In [1][2] and recently under general assumptions in [3], it has been shown that the capacity region is given by

$$\Gamma = \sum_{\mathbf{s} \in \mathcal{S}} \pi(\mathbf{s}) \text{ Convex-Hull}\{\mathbf{D}(\mathbf{s}, \mathbf{I}) | \mathbf{I} \in \mathcal{I}\}. \qquad (7)$$

Moreover, it has been shown that the GMWM policy is throughput optimal in that it stabilizes the network for all input rates that are strictly inside $\Gamma$ [1][2][3]. This policy at each time-slot sets $\mathbf{X} = \mathbf{X}(t)$ and $\mathbf{s} = \mathbf{s}(t)$, and uses the

---

schedule vector $\mathbf{I}^*(t)$ that is *argmax*[3] to the following GMWM optimization problem:

$$\max \sum_{l=1}^{N} X_l D_l(\mathbf{s}, \mathbf{I}), \qquad \text{subject to } \mathbf{I} \in \mathcal{I}. \qquad (8)$$

Hence, the GMWM policy uses a schedule vector that maximizes the sum of backlog-rate products. However, note that the optimization problem given in (8) can be NP-hard [8][4]. We therefore consider a policy based on randomized algorithms that can provide approximate solutions to the optimization problem in (8). In the following, we first elaborate on the structure of the considered sub-optimal algorithms.

### A. Sub-optimal Algorithms Approximating GMWM Problem

In this paper, we assume that there exists a randomized algorithm $A$, either centralized or distributed, that at each time-slot $t$ provides the network with a *candidate* schedule $\mathbf{I}^r(t)$ from the set $\mathcal{I}$. We use superscript $r$ to emphasize that $\mathbf{I}^r(t)$ is a candidate schedule vector selected by the randomized algorithm $A$. In general, the distribution of the selected schedule $\mathbf{I}^r(t)$ depends on $\mathbf{X}(t)$ and $\mathbf{s}(t)$. We define $\mu_{\mathbf{X}(t),\mathbf{s}(t)}(\cdot)$ to be the distribution for $\mathbf{I}^r(t)$. Note that the policy developed in this paper does not need the knowledge of $\mu_{\mathbf{X}(t),\mathbf{s}(t)}(\cdot)$, and only requires the algorithm $A$ to satisfy Properties 1 and 2, as will be discussed shortly.

Given $\mathbf{X}$ and $\mathbf{s}$, let the *optimal schedule* and the *optimal rate* associated with the GMWM problem be $\mathbf{I}^*(\mathbf{X}, \mathbf{s})$ and $\mathbf{D}^*(\mathbf{X}, \mathbf{s})$, where

$$\mathbf{I}^*(\mathbf{X}, \mathbf{s}) = \operatorname*{argmax}_{\mathbf{I} \in \mathcal{I}} \mathbf{X} \mathbf{D}(\mathbf{s}, \mathbf{I}), \qquad (9)$$

and

$$\mathbf{D}^*(\mathbf{X}, \mathbf{s}) = \mathbf{D}(\mathbf{s}, \mathbf{I}^*(\mathbf{X}, \mathbf{s})). \qquad (10)$$

Note that in the above, $\mathbf{X} \mathbf{D}(\mathbf{s}, \mathbf{I})$ denotes the scalar product of the vectors $\mathbf{X}$ and $\mathbf{D}(\mathbf{s}, \mathbf{I})$, and for simplicity, we have dropped the dot operator. In the rest of the paper, we use the same method to denote scalar products of vectors. In addition, note that by (1), $\mathbf{D}(\mathbf{s}, \mathbf{I})$ is the rate vector corresponding to the channel state $\mathbf{s}$ and the schedule vector $\mathbf{I}$. We assume the algorithm $A$ has the following property.

**Property 1.** *There exist a constant $\zeta$, $0 \leq \zeta < 1$, and a constant $\delta$, $\delta > 0$, such that, for any given $\mathbf{X}$ and $\mathbf{s} \in \mathcal{S}$, with probability at least $\delta$, the algorithm $A$ finds a candidate vector $\mathbf{I}^r$ that satisfies the following:*

$$\mathbf{X} \mathbf{D}(\mathbf{s}, \mathbf{I}^r) \geq (1 - \zeta) \mathbf{X} \mathbf{D}^*(\mathbf{X}, \mathbf{s}). \qquad (11)$$

This property simply states that the selected schedule $\mathbf{I}^r$ with probability at least $\delta$ is within $\zeta$-neighborhood of the optimal solution in terms of the backlog-rate product. We can consider this property as a generalized version of the ones in [5][6][8][7][9], modeling the sub-optimality of the algorithm $A$ through the introduction of the pair $(\zeta, \delta)$. The following further details the structure of the algorithm $A$ by stating a property for the distribution set $\{\mu_{\mathbf{X},\mathbf{s}}(\cdot); \mathbf{s} \in \mathcal{S}, \mathbf{X} \in (\{0\} \cup \mathbb{Z}^+)^N\}$.

**Property 2.** *Consider two queue-length vectors $\mathbf{X}_1$ and $\mathbf{X}_2$, and suppose $\|\mathbf{X}_2 - \mathbf{X}_1\| < C$ for a given constant $C > 0$. For*

any given positive $\epsilon$, there exists a sufficiently large $B_{1,\epsilon}^C$ such that if $\|\mathbf{X}_1\| \geq B_{1,\epsilon}^C$, then

$$\sum_{\mathbf{I} \in \mathcal{I}} |\mu_{\mathbf{X}_2,\mathbf{s}}(\mathbf{I}^r = \mathbf{I}) - \mu_{\mathbf{X}_1,\mathbf{s}}(\mathbf{I}^r = \mathbf{I})| < \epsilon,$$

for all $\mathbf{s} \in \mathcal{S}$. Moreover, for any two given $\mathbf{X}_1$ and $\mathbf{X}_2$, if $\mathbf{X}_2 = \beta \mathbf{X}_1$, for some $\beta > 0$, then, for all $\mathbf{s} \in \mathcal{S}$ and $\mathbf{I} \in \mathcal{I}$,

$$\mu_{\mathbf{X}_2,\mathbf{s}}(\mathbf{I}^r = \mathbf{I}) = \mu_{\mathbf{X}_1,\mathbf{s}}(\mathbf{I}^r = \mathbf{I}).$$

This property states that the distributions for $\mathbf{I}^r$ are almost the same when two queue-length vectors are *close* and *large*. This property also states that the distribution is exactly the same if two queue-length vectors differ only by a multiplicative scalar factor. These statements may naturally hold since the objective function in (8) is a continuous function of $\mathbf{X}$, and assuming $\|\mathbf{X}_2 - \mathbf{X}_1\| < C$, for any $\mathbf{s}$ and $\mathbf{I}$, by (2), we must have

$$|\mathbf{X}_2 \mathbf{D}(\mathbf{s}, \mathbf{I}) - \mathbf{X}_1 \mathbf{D}(\mathbf{s}, \mathbf{I})| < \sqrt{N} C D_{max}.$$

Hence, finite changes in the queue-length vector have a finite impact on the backlog-rate product. This and the fact that $\mathbf{X} \mathbf{D}^*(\mathbf{X}, \mathbf{s})$ linearly[4] increases with $\|\mathbf{X}\|$ suggest that the impact when normalized to $\mathbf{X}_1 \mathbf{D}^*(\mathbf{X}_1, \mathbf{s})$ can be arbitrarily small if $\|\mathbf{X}_1\|$ is sufficiently large. We therefore expect the algorithm $A$, for each pair $(\mathbf{s}, \mathbf{I})$, to *see* similar normalized values[5] of the backlog-rate products corresponding to $\mathbf{X}_1$ and $\mathbf{X}_2$. Hence, for given $\mathbf{X}_1$, $\mathbf{X}_2$, and $\mathbf{s}$, the algorithm is expected to assign similar probabilities for $\mathbf{I}^r = \mathbf{I}$ when $\|\mathbf{X}_2 - \mathbf{X}_1\| < C$, and $\|\mathbf{X}_1\|$ is sufficiently large. In the case where $\mathbf{X}_2 = \beta \mathbf{X}_1$, the backlog-rate product corresponding to $\mathbf{X}_2$, for all $\mathbf{s} \in \mathcal{S}$ and $\mathbf{I} \in \mathcal{I}$, is a $\beta$-scaled version of the one for $\mathbf{X}_1$. Therefore, we expect the distribution for $\mathbf{I}^r$ corresponding to $\mathbf{X}_2$ to be exactly the same as the one corresponding to $\mathbf{X}_1$. Having detailed the structure of the algorithm $A$, we next focus on the operation of LM-RSP.

### B. LM-RSP's Operation and Scheduling

We start by defining several useful functions. Let $\mathbf{D}^r(t)$ and $\mathbf{D}'(t-1)$ be defined as

$$\mathbf{D}^r(t) = \mathbf{D}(\mathbf{s}(t), \mathbf{I}^r(t)),$$

and

$$\mathbf{D}'(t-1) = \mathbf{D}(\mathbf{s}(t), \mathbf{I}(t-1)), \qquad (12)$$

respectively. According to the above definitions, we see that if the network uses the candidate schedule $\mathbf{I}^r(t)$ at time-slot $t$, the resulting rate vector will be $\mathbf{D}^r(t)$ whereas if the network keeps using the schedule vector of the previous time-slot, $\mathbf{I}(t-1)$, $\mathbf{D}'(t-1)$ will be the rate vector (at time-slot $t$). In addition, let $\varphi(t)$ be defined as

$$\varphi(t) = \frac{\mathbf{X}(t)(\mathbf{D}^r(t) - \mathbf{D}'(t-1))}{\max(\mathbf{X}(t)\mathbf{D}^r(t), \mathbf{X}(t)\mathbf{D}'(t-1)) + \alpha\|\mathbf{X}(t)\|}, \qquad (13)$$

where $\alpha$ is a *positive* but otherwise *arbitrary* constant. Later in this section, we elaborate on the motivation to consider a non-zero value for $\alpha$. Based on the above definition, $\varphi(t)$ measures the *normalized*[6] improvement in terms of backlog-rate product when the candidate schedule $\mathbf{I}^r(t)$ is preferred over $\mathbf{I}(t-1)$ at time-slot $t$. We define $\varphi(t) = 0$ if $\mathbf{X}(t) = 0$.

---

[3]If there are more than one schedule vector maximizing the summation in (8), we define the argmax to be any of such schedule vectors.

[4]Later, we show that $\mathbf{X} \mathbf{D}^*(\mathbf{X}, \mathbf{s}) \geq \frac{\nu}{\sqrt{N}}\|\mathbf{X}\|$, where $\nu$ is a positive system dependent constant as defined in (19).

[5]Here, for a given $\mathbf{X}$, the product is normalized to $\mathbf{X} \mathbf{D}^*(\mathbf{X}, \mathbf{s})$.

[6]Here, by normalization, we mean division by a well-defined function.

We are now ready to describe the operation of LM-RSP. We assume that the policy, either centralized or distributed, takes as the input the vectors $\mathbf{I}(t-1)$, $\mathbf{X}(t)$, and $\mathbf{s}(t)$; and using the algorithm $A$, updates the schedule vector $\mathbf{I}(t)$ according to the following:

- Using the algorithm $A$, the policy selects $\mathbf{I}(0)$ according to the initial $\mathbf{X}(0)$ and $\mathbf{s}(0)$.
- For $t > 0$, it determines $\mathbf{I}(t)$ through the following steps.
  - First, the policy uses the algorithm $A$ to select $\mathbf{I}^r(t)$ according to $\mathbf{X}(t)$ and $\mathbf{s}(t)$.
  - Next, it updates $\mathbf{I}(t)$ according to the following rule:
  
  $$\mathbf{I}(t) = \begin{cases} \mathbf{I}^r(t) & \text{with probability } f(\varphi(t)) \\ \mathbf{I}(t-1) & \text{otherwise} \end{cases}$$
  
  where $\varphi(t)$ is defined in (13), and $f : (-1,1) \to [0,1]$ is a non-decreasing continuous function.

We assume the $f(\varphi) - 0.5$ is an odd function of $\varphi$, and $f(\varphi)$ has the property that

$$f(\varphi) = \begin{cases} 1, & \varphi \geq \rho \\ 0, & \varphi \leq -\rho \end{cases},$$

where $0 < \rho < 1$. In the rest of the paper, we assume the function $f(\varphi)$ is linear in the range $[-\rho, \rho]$ and $f(\varphi) = 0.5 + \frac{\varphi}{2\rho}$ for $|\varphi| < \rho$; we leave finding the optimal $f(\varphi)$ as an interesting open problem for future research. Considering the definition of $\varphi(t)$ and the properties for $f(\cdot)$, it is easy to see that w.p. 1

$$\mathbf{X}(t)(\mathbf{D}(t) - (1 - \rho)\mathbf{D}'(t-1)) > -\rho\alpha\|\mathbf{X}(t)\|. \quad (14)$$

Note that a similar inequality holds by replacing $\mathbf{D}'(t-1)$ with $\mathbf{D}^r(t)$.

The above description suggests that the memory requirement of the policy is *linear* in the number of data flows $N$. This is because the only past information required to update $\mathbf{I}(t)$ in each time-slot is the vector $\mathbf{I}(t-1)$ whose size is proportional to $N$. As mentioned earlier, other similar proposals in the context of time-varying channels [6][7], store one vector for *each* possible channel state. However, the number of states increases exponentially with $N$, which implies that these proposals require an exponentially increasing memory storage.

One subtle point in the design of the policy is the introduction of $\varphi(t)$ and $f(\varphi(t))$ in the update process of $\mathbf{I}(t)$. These functions allow LM-RSP to take soft decisions when comparing two different schedules, generalizing similar previous approaches in [5][6][7][9]. Specifically, these functions enable LM-RSP to probabilistically choose either of the schedules, $\mathbf{I}(t-1)$ or $\mathbf{I}^r(t)$, according to the value of $\varphi(t)$; a larger positive $\varphi(t)$ implies that selecting $\mathbf{I}^r(t)$ as the schedule vector results in a larger backlog-rate product, which according to the monotonicity of $f(\varphi)$, increases the probability of selecting $\mathbf{I}^r(t)$. Similarly, a smaller negative $\varphi(t)$ increases the chance of selecting $\mathbf{I}(t-1)$ as the schedule vector for time-slot $t$.

However, a mere generalization is not the main motivation for the introduction and use of $\varphi(t)$ and $f(\varphi(t))$. The main motivation is to make the distribution of $\mathbf{I}(t)$ *continuous* with respect to $\mathbf{X}(t)$ for large $\|\mathbf{X}(t)\|$, and thus, is analysis-inspired. More specifically, we have the following fact, which results from the continuity of $f(\cdot)$, the assumption that $\alpha > 0$, and the fact that the distribution of $\mathbf{I}(t)$ is completely determined by the values for $\mathbf{I}(t-1)$, $\mathbf{X}(t)$, $\mathbf{I}^r(t)$, and $\mathbf{s}(t)$.

**Fact 2.** *Suppose two vectors $\mathbf{X}_1$ and $\mathbf{X}_2$ are given such that $\|\mathbf{X}_2 - \mathbf{X}_1\| < C$. For any given positive $\epsilon$, there exists a sufficiently large $B_{3,\epsilon}^C$ such that if $\|\mathbf{X}_1\| \geq B_{3,\epsilon}^C$, then for all $\mathbf{I}^r(t) \in \mathcal{I}$, $\mathbf{I}(t-1) \in \mathcal{I}$, $\mathbf{s}(t) \in \mathcal{S}$, and $\mathbf{I} \in \{\mathbf{I}^r(t), \mathbf{I}(t-1)\}$ the following holds:*

$$|p(\mathbf{I}(t) = \mathbf{I}|\mathbf{X}(t) = \mathbf{X}_2, \mathbf{s}(t), \mathbf{I}^r(t), \mathbf{I}(t-1))$$
$$- p(\mathbf{I}(t) = \mathbf{I}|\mathbf{X}(t) = \mathbf{X}_1, \mathbf{s}(t), \mathbf{I}^r(t), \mathbf{I}(t-1))| < \epsilon. \quad (15)$$

One important point that should not be buried under the main motivation is that introducing $\varphi(t)$ and $f(\varphi)$ allows us to consider and embed in our model the efficiency of the update rule when it is implemented distributedly. The update rule in its original form [5][6][7] simply takes a decision based on only the *sign* of $\varphi(t)$ and uses the schedule vector, either $\mathbf{I}(t-1)$ or $\mathbf{I}^r(t)$, with the larger backlog-rate product. When the update rule is implemented in a distributed manner, however, the estimates of $\mathbf{X}(t)$, $\mathbf{X}(t)\mathbf{D}^r(t)$, and $\mathbf{X}(t)\mathbf{D}'(t-1)$ can be erroneous. These inaccurate estimates may make the computed $\varphi(t)$ and, in particular, its sign to be different from their actual values. Hence, the sign of $\varphi(t)$ alone may not be sufficient for comparison purposes. This suggests to use the *value*, and not only the sign, of $\varphi(t)$ to evaluate and select the vectors $\mathbf{I}(t-1)$ and $\mathbf{I}^r(t)$. In this case, while the choice for a positive $\alpha$ is *arbitrary*, the value of $\rho$ may be adjusted to account for the extent within which the estimates are inaccurate. Note that when LM-RSP is implemented in a centralized manner, the choice for both $\alpha$ and $\rho$ is *arbitrary* as long as $\alpha > 0$ and $0 < \rho < 1$.

### C. Complexity of LM-RSP

As mentioned earlier, at each time-slot $t$, the GMWM policy solves the optimization problem in (8) to find $\mathbf{I}^*(\mathbf{X}(t), \mathbf{s}(t))$ as the schedule vector. This problem is in general non-convex due to physical layer interferences, and can be NP-hard [4]. In contrast, LM-RSP assumes access to the algorithm $A$ whose complexity depends on the value of the pair $(\zeta, \delta)$. For instance, when $\delta = 1$ and $\zeta = 0$, the algorithm $A$ always returns the optimal solution, and when $\delta = |\mathcal{I}|^{-1}$ and $\zeta = 0$, the algorithm simply selects schedule vectors with equal probabilities. It is easy to see that the latter case, with purely random selection, can make the complexity of LM-RSP linear in $N$. This special case is attractive mainly from a theoretical point of view since it achieves throughput-optimality in the limit of highly correlated channels, despite an exponentially increasing delay due to the exponentially decreasing $\delta$ with the number of data flows (see Corollaries 2 and 6). More interesting examples are discussed in [30], where a distributed algorithm is developed with a time-complexity of $x \cdot c \cdot \epsilon^{-1} \log \epsilon^{-1} \log N$, where $c$ is a constant, and $x \geq 1$ and $\epsilon > 0$ are tuning parameters such that $\zeta = 1 - \frac{1}{4+\epsilon}$ and $\delta = 1 - N^{1-x}$.

## V. LM-RSP STABILITY REGION

In this section, we study the stability region of the network under LM-RSP by first providing several key definitions.

### A. Key Definitions

For notational convenience, in the rest of the paper, where appropriate, we use subscripts also to show dependencies on time; hence, e.g., we have $\mathbf{X}_t \triangleq \mathbf{X}(t)$. Let $\Upsilon(\mathbf{X}_t)$ be defined as

$$\Upsilon(\mathbf{X}_t) = \mathbb{E}[\mathbf{X}_t \mathbf{D}^*(\mathbf{X}_t, \mathbf{s})], \quad (16)$$

where the expectation is over the steady-state distribution of the channel process, and $\mathbf{D}^*(\mathbf{X}_t, \mathbf{s})$ is defined in (10). Based on this definition, $\Upsilon(\mathbf{X}_t)$ denotes the expected value of the *maximum* backlog-rate product and, thus, is the expected backlog-rate product if the GMWM policy is used, and the queue-length vector is fixed at $\mathbf{X} = \mathbf{X}_t$. The quantity $\Upsilon(\mathbf{X}_t)$, therefore, can serve as a *benchmark* to measure the performance of suboptimal policies.

In our analysis, we often encounter distributions and expected values of random variables where after a particular time $t$, queue dynamics are ignored. To make this notationally clear, suppose a r.v. $Z$ is given, which can be a function of the channel process and the selected schedules. We define $\bar{\mathbb{E}}_{\mathbf{X}_t}[Z]$ and $\bar{p}_{\mathbf{X}_t}(Z = z)$ as, respectively, the expectation of the r.v. $Z$ and the probability that $Z = z$, given the hypothesis that at any time $t'$, where $t' > t$, the policy updates $\mathbf{I}(t')$ by assuming $\mathbf{X}(t') = \mathbf{X}(t)$. In other words, these notations emphasize that after time $t$, the policy makes decisions based on the *old* queue-length information at time $t$. These notations further assume that the updated channel state information $\mathbf{s}(t'), t' > t$, is available. Note that without the above hypothesis, both updated queue and channel state information are available for LM-RSP to update the schedule vector.

Having introduced $\bar{\mathbb{E}}_{\mathbf{X}_t}[\cdot]$, we define $\Psi_{\mathbf{Y}_t}^K$ as

$$\Psi_{\mathbf{Y}_t}^K = \frac{\sum_{m=0}^{K-2} \bar{\mathbb{E}}_{\mathbf{X}_t}[\mathbf{X}_t(\mathbf{D}_{t+m} - (1-\rho)\mathbf{D}'_{t+m})|\mathbf{Y}_t]}{K \Upsilon(\mathbf{X}_t)}. \quad (17)$$

By the definition in (12), $\mathbf{D}'_t$ is the rate vector corresponding to the schedule vector in the current time-slot, $\mathbf{I}(t)$, and the channel state in the next time-slot, $\mathbf{s}(t+1)$, implying that $\mathbf{D}_{t+m} - (1-\rho)\mathbf{D}'_{t+m}$ for small values of $\rho$ approximately shows the changes in the rate vector $\mathbf{D}_{t+m}$ due to channel variations. Note that the sequence of $\mathbf{D}_{t+m}$'s not only depends on a particular realization of the channel states but also on the randomized algorithm $A$ that finds the candidate schedules. Therefore, when $K$ is large, $\Psi_{\mathbf{Y}_t}^K$ measures the relative changes in the backlog-rate products, due to channel variations, over a long horizon, while implicitly embedding the effects of the algorithm $A$. This implies that $\Psi_{\mathbf{Y}_t}^K$ can be used as a measure for the channel correlation since $\Psi_{\mathbf{Y}_t}^K$ becomes small for small values of $\rho$ if the channel states are highly correlated. It is important to note that in the definition of $\Psi_{\mathbf{Y}_t}^K$, the expectation is of the type $\bar{\mathbb{E}}_{\mathbf{X}_t}[\cdot]$, and hence, queue variations after time $t$ do not affect $\Psi_{\mathbf{Y}_t}^K$. In addition, note that $\Psi_{\mathbf{Y}_t}^K$ not only depends on $K$ and $\mathbf{Y}_t$ but also implicitly depends on $t$.

Similarly, let

$$\Phi_{\mathbf{Y}_t}^K = \frac{\sum_{m=0}^{K-1} \bar{\mathbb{E}}_{\mathbf{X}_t}[\mathbf{X}_t \mathbf{D}_{t+m}|\mathbf{Y}_t]}{K \Upsilon(\mathbf{X}_t)}. \quad (18)$$

This definition introduces $\Phi_{\mathbf{Y}_t}^K$ as the time average of backlog-rate product normalized to the benchmark $\Upsilon(\mathbf{X}_t)$. Hence, we can use $\Phi_{\mathbf{Y}_t}^K$ as a measure to compare the performance of LM-RSP with that of the GMWM policy.

As for one other definition, let

$$\nu = \min_{1 \le l \le N} \sum_{\mathbf{s} \in \mathcal{S}} \pi(\mathbf{s}) \max_{\mathbf{I} \in \mathcal{I}} D_l(\mathbf{s}, \mathbf{I}). \quad (19)$$

Thus, $\nu$ is the minimum of the average maximum transmission rate for *individual* links, over all links in the network. This parameter is a fundamental property of the system. It immediately follows that if for every link, there is at least one state in which the transmission rate is non-zero, then $\nu > 0$, which is assumed throughout the paper. One importance of this parameter is that we can obtain a lower bound for $\Upsilon(\mathbf{X}_t)$:

$$\Upsilon(\mathbf{X}_t) \ge \nu \max_{1 \le l \le N} X_l(t) \ge \frac{\nu}{\sqrt{N}} \|\mathbf{X}_t\|. \quad (20)$$

Now, we define a key parameter $\theta$ that represents the fraction of the capacity region $\Gamma$ that can be stabilized by LM-RSP. Specifically, we define $\theta$ as

$$\theta = \liminf_{K \to \infty} \inf_{\mathbf{Y}_t} \max(1 - \zeta' - \frac{1-\delta}{\delta}\Psi_{\mathbf{Y}_t}^K - \frac{\sqrt{N}\rho\alpha}{\delta\nu}, \Phi_{\mathbf{Y}_t}^K), (21)$$

where

$$\zeta' = (1 - (1-\rho)(1-\zeta)). \quad (22)$$

We assume $\frac{\sqrt{N}\rho\alpha}{\delta\nu} < 1$ or, in other words,

$$\rho\alpha < \frac{\delta\nu}{\sqrt{N}}. \quad (23)$$

This assumption can be, in general, a necessary condition for the positivity of the first argument in the $\max$ operator of (21). By the definition of the $\liminf$, we have the following fact.

**Fact 3.** *For any positive $\epsilon > 0$, there exists a sufficiently large $K_\epsilon^{(\theta)}$, such that for all $\mathbf{Y}_t$ and $K \ge K_\epsilon^{(\theta)}$, the following holds:*

$$\max(1 - \zeta' - \frac{1-\delta}{\delta}\Psi_{\mathbf{Y}_t}^K - \frac{\sqrt{N}\rho\alpha}{\delta\nu}, \Phi_{\mathbf{Y}_t}^K) > \theta - \epsilon.$$

To shed light on the properties of the parameter $\theta$, and also to consider an important special case, suppose the channel process is Markovian. According to our assumptions, there is a finite number of channel states and schedules. Therefore, given the hypothetical condition that the queue-length vector is frozen at $\mathbf{X}_t$ after time $t$, as assumed in the definition of $\Psi_{\mathbf{Y}_t}^K$ and $\Phi_{\mathbf{Y}_t}^K$, the joint process of rates and the channel states will be a Markov chain with a finite number of states. If the joint process has a single communicating class and is aperiodic, e.g., when the randomized algorithm selects any schedule with a positive probability, and the channel process is irreducible and aperiodic, then the joint process will be positive recurrent and will have a steady-state distribution. Hence, as $m \to \infty$, $\mathbf{D}_{t+m}$ weakly converges to a random vector $\mathbf{D}$ whose distribution depends on the channel distribution, the algorithm $A$, the update rule, and the given vector $\mathbf{X} = \mathbf{X}_t$. A similar discussion also holds for $\mathbf{D}'_{t+m}$. Therefore, in the limit of large $K$, both $\Psi_{\mathbf{Y}_t}^K$ and $\Phi_{\mathbf{Y}_t}^K$ become independent of initial $\mathbf{I}(t)$ and $\mathbf{s}(t)$. In particular, assuming $\mathbf{X} = \mathbf{X}_t$, as $K \to \infty$, we have

$$\Psi_{\mathbf{Y}_t}^K \to \Psi_{\mathbf{X}}^\infty \triangleq \Psi_{\mathbf{Y}_t}^\infty = \frac{\mathbb{E}[\mathbf{X}(\mathbf{D} - (1-\rho)\mathbf{D}')]}{\Upsilon(\mathbf{X})}, \quad (24)$$

and

$$\Phi_{\mathbf{Y}_t}^K \to \Phi_{\mathbf{X}}^\infty \triangleq \Phi_{\mathbf{Y}_t}^\infty = \frac{\mathbb{E}[\mathbf{X}\mathbf{D}]}{\Upsilon(\mathbf{X})}, \quad (25)$$

where in the above expressions, expectations are taken with respect to the distributions for $\mathbf{D}$ and $\mathbf{D}'$, and by using $\Psi_{\mathbf{Y}_t}^\infty$ and $\Psi_{\mathbf{X}}^\infty$, we have misused the notation to emphasize that $\Psi_{\mathbf{Y}_t}^\infty$ and $\Phi_{\mathbf{Y}_t}^\infty$ depend on $\mathbf{Y}_t$ only through $\mathbf{X} = \mathbf{X}_t$.

As one other observation for this special case, note that by the Markovian nature of the channel process, $\Psi_{\mathbf{Y}_t}^K$ and $\Phi_{\mathbf{Y}_t}^K$ become independent of $t$ when $\mathbf{Y}_t$ is given. This independence, Property 2, and the update rule further imply that the distributions for $\Psi_{\mathbf{Y}_t}^K$ and $\Phi_{\mathbf{Y}_t}^K$ do not depend on $\|\mathbf{X}_t\|$. Therefore, in this case, despite the fact that the vector $\mathbf{X}_t$ is discrete, for the purpose of taking the inf over $\mathbf{Y}_t$ in the definition of $\theta$, we can replace $\mathbf{X}_t$, as one element of $\mathbf{Y}_t$, with $\frac{\mathbf{X}_t}{\|\mathbf{X}_t\|}$. In addition, since the inf is taken over all possible $t$'s, $\frac{\mathbf{X}_t}{\|\mathbf{X}_t\|}$ can take all

possible directions and can be any unit vector in the limit of large $t$. Hence, by (24) and (25), for Markovian channels, we must have

$$\theta = \inf_{\mathbf{X}:\|\mathbf{X}\|=1} \max(1 - \zeta' - \frac{1-\delta}{\delta}\Psi_{\mathbf{X}}^\infty - \frac{\sqrt{N}\rho\alpha}{\delta\nu}, \Phi_{\mathbf{X}}^\infty). \quad (26)$$

### B. Theorem on Stability Region

The following is the main result of this paper on the stability region of LM-RSP.

**Theorem 1.** *Suppose the mean arrival rate vector* $\mathbf{a}$ *lies strictly inside* $\theta\Gamma$, *where* $\theta$ *is defined in (21), and* $\theta\Gamma$ *is a region that contains* $\theta$-*scaled of all rates in* $\Gamma$, *i.e.,* $\theta\Gamma = \{\mathbf{a}_\theta | \exists \mathbf{a} \in \Gamma : \mathbf{a}_\theta = \theta\mathbf{a}\}$. *We have the following:*

(a) *There exist non-negative constants* $\beta_{\mathbf{s},\mathbf{I}}$'s *such that*

$$\mathbf{a} = \sum_{\mathbf{s}\in\mathcal{S}} \pi(\mathbf{s}) \sum_{\mathbf{I}\in\mathcal{I}} \beta_{\mathbf{s},\mathbf{I}} \mathbf{D}(\mathbf{s},\mathbf{I}), \quad (27)$$

*and*

$$\epsilon \triangleq \theta - \max_{\mathbf{s}\in\mathcal{S}} \sum_{\mathbf{I}\in\mathcal{I}} \beta_{\mathbf{s},\mathbf{I}} > 0.$$

(b) *Under LM-RSP, the system described in Section III is stable in the mean, i.e.,*

$$\limsup_{T\to\infty} \frac{1}{T+1} \sum_{t=0}^{T} \mathbb{E}\big[\|\mathbf{X}_t\|\big] < \infty.$$

*Proof.* The proof of the theorem is given in the appendix. $\square$

### C. Insights into the Region $\theta\Gamma$

Here, we discuss several practical implications of the above theorem by first focusing on general channel processes and then considering an example of Markovian channel states.

*1) Insights Assuming General Channel Processes:* As the first point, the theorem suggests that a scaled version of the capacity region $\Gamma$ can be supported by LM-RSP. The theorem, moreover, shows that the scaling factor is $\theta$, which by definition depends on the limiting behavior of the policy when queue dynamics are ignored. Recall that $\theta$ is a function of $\Psi_{\mathbf{Y}_t}^K$ and $\Phi_{\mathbf{Y}_t}^K$. As explained in Section V-A, $\Psi_{\mathbf{Y}_t}^K$ measures normalized rate changes due to channel variations over time when queue variations are ignored after time $t$. Since, for a given timeslot, the policy updates the schedule vector by comparing a candidate schedule with the one used in its previous time-slot, we expect that large channel variations, and thus, large $\Psi_{\mathbf{Y}_t}^K$, negatively affect the update process, and hence, $\theta$. On the other hand, as explained earlier, for a given $\mathbf{Y}_t$, $\Phi_{\mathbf{Y}_t}^K$ is a measure to compare LM-RSP with the GMWM policy; a larger and close-to-one value for $\Phi_{\mathbf{Y}_t}^K$ indicates that LM-RSP uses schedule vectors with similar backlog-rate products to the ones resulting from the GMWM policy, and a smaller and close-to-zero value for $\Phi_{\mathbf{Y}_t}^K$ indicates that LM-RSP is performing poorly compared to the GMWM policy. As a result, we intuitively expect $\theta$ to be as large as the least value of $\Phi_{\mathbf{Y}_t}^K$ for large $K$'s, which is the inf (over $\mathbf{Y}_t$) of $\Psi_{\mathbf{Y}_t}^K$ for large $K$'s. The expression for $\theta$ exactly reflects these observations.

We also observe that the parameters $\alpha$ and $\rho$ can directly affect $\theta$ through the term $\frac{\sqrt{N}\rho\alpha}{\delta\nu}$ and indirectly through the terms $\Psi_{\mathbf{Y}_t}^K$ and $\Phi_{\mathbf{Y}_t}^K$. Recall that these two parameters must be positive for continuity purposes but, otherwise, can be chosen arbitrarily[7]. Note that $\delta$ is a given parameter, and $\nu$ can be estimated readily. Hence, we might naturally try to choose $\alpha$ and $\rho$ such that $\frac{\sqrt{N}\rho\alpha}{\delta\nu}$ is arbitrarily small. In fact, assuming $\theta_{lim}$ exists, where

$$\theta_{lim} = \lim_{\alpha,\rho\to 0} \theta, \quad (28)$$

we can ensure $\theta\Gamma$ contains a region arbitrarily close to $\theta_{lim}\Gamma$ by assuming sufficiently small values for $\rho$ and $\alpha$, which gives rise to the following corollary. However, note that, as shown later, the delay bound can increase proportionally with $\frac{1}{\rho\alpha}$.

**Corollary 1.** *For any input rate strictly inside the region* $\theta_{lim}\Gamma$, *the parameters* $\alpha$ *and* $\rho$ *can be chosen sufficiently small such that the system described in Section III is stable under LM-RSP.*

We now consider the effect of channel variations on $\theta$. Suppose channel states are highly correlated. This implies that $\mathbf{D}'_{t+m} \simeq \mathbf{D}_{t+m}$. Since by definition, $\mathbb{E}[\mathbf{XD}] \simeq \Upsilon(\mathbf{X})$, from (17), we have $\Psi_{\mathbf{Y}_t}^K \leq \rho'$, where $\rho' \simeq \rho$. Assuming that $\rho$ and $\alpha$ are sufficiently small and using (21), we have that $\theta \geq 1 - \zeta' \simeq 1 - \zeta$. It is interesting to see how the presence of the term $\frac{1-\delta}{\delta}$ in $\theta$ is canceled by the channel correlation. Note that the term $\frac{1-\delta}{\delta}$ is the average number of times that the algorithm $A$ must be run before (11) holds for a fixed $\mathbf{X}$ and $\mathbf{s}$. The effect of this term is reduced when channel correlation is high, which manifests itself in a small $\Psi_{\mathbf{Y}_t}^K$. We can also easily prove that if the candidate schedule returned by the algorithm $A$ is used *without* any comparison in each timeslot, in general, the scaling factor becomes $\delta(1-\zeta)$. Therefore, we see that LM-RSP improves the capacity region scaling from $\delta(1-\zeta)$ to at least $1-\zeta$ and, exploiting channel correlations, reduces the uncertainty of the randomized algorithm $A$ in selecting a candidate schedule satisfying (11). A special case is where $\zeta = 0$, which implies $\theta \geq 1$ in the limit of $\rho \to 0$, and thus, $\theta = 1$. Since $\theta = 1$ means throughput optimality, we conclude that simple linear-complexity algorithms, see the discussion in Section IV-C, are sufficient to attain throughput optimality arbitrarily closely, reminiscent of the results in [5]. We summarize the above in the following corollary.

**Corollary 2.** *The stability region* $\theta\Gamma$ *contains the region* $(1-\zeta)\Gamma$ *and* $\theta_{lim} \geq 1-\zeta$ *in the limit of highly correlated channel states and small* $\rho$ *and* $\alpha$. *In particular, when* $\zeta = 0$, *in the limit, the region* $\theta\Gamma$ *expands to the capacity region* $\Gamma$, *and LM-RSP becomes throughput optimal.*

*2) Insights Assuming Markovian Channel Processes:* Our discussion so far considers networks with general channel processes. In the following, to obtain specific results, we focus on an important class of Markovian channels and well-investigated interference models. Suppose the channel states of wireless links are independent. Furthermore, suppose the state of each link is a Markov chain with two states, namely the $g$ state representing the "good" state and the $b$ state representing the "bad" state. We assume that the state of a link in each transition can take a different value with probability $r$. Hence, $r$ may represent *individual* link variation rate over one time-slot. As the worst-case scenario, we assume in the $b$ state the transmission rate is zero. We do not impose any assumption, other than positivity, on the transmission rate in the $g$ state. Therefore, when two links are in their $g$ states, they can see possibly different but non-zero transmission rates.

---

[7]As discussed in Section IV-B, if the update rule is implemented distributedly, $\rho$ may be used to model the inefficiencies in implementing LM-RSP. In this case, the choice for a positive $\alpha$ is still arbitrary.

As for the interference, we consider the classic *node-exclusive interference model* [8][4][9][10], where a node can only send to or receive from one other node at any time. This interference model motivates us to view the network as a graph $G(V, E)$, where $V$ is the set of users and $E$ is the set of all links in the network. Given this graph, a *valid* schedule is a *matching*, where a matching is a set of edges no two of which share a common vertex. We assume the algorithm $A$ always returns a matching with respect to $G$, ensuring that the schedule vector $\mathbf{I}_t$ is also a matching. Note that our discussion here easily extends to the more general $\kappa$-hop interference model [4], according to which, no two links within $\kappa$ hops can successfully transmit at the same time.

Having defined the channel and interference models, we now derive an upper-bound for $\Psi_{\mathbf{Y}_t}^K$. Recall that $\Psi_{\mathbf{Y}_t}^K$ is almost the time average of $\bar{\mathbb{E}}_{\mathbf{X}_t}[\mathbf{X}_t(\mathbf{D}_{t+m} - (1-\rho)\mathbf{D}'_{t+m})|\mathbf{Y}_t]$ taken over $m$, $0 \leq m \leq K - 2$, and normalized to $\Upsilon(\mathbf{X}_t)$. Consider the time $t + m$, and suppose the $l_{\text{th}}$ link is in its $g$ state and is scheduled to receive non-zero transmission rate. In the next time-slot, with probability $r$, the state of this link changes to the $b$ state, whose definition implies that $D'_l(t + m) = 0$. On the other hand, with probability $1 - r$, the link stays in its $g$ state. Since by definition, $\mathbf{D}'_{t+m}$ is the rate at time $t + m + 1$ but with the schedule vector $\mathbf{I}_{t+m}$ used at time $t+m$, and since by assumption schedule vectors are matchings, we see that no links that can possibly interfere with the $l_{th}$ link are scheduled at time $t+m+1$. Hence, when the $l_{th}$ link stays in its $g$ state, we must have $D'_l(t+m) = D_l(t+m)$. In the case where the $l_{th}$ link is in its $b$ state at time $t+m$, or not scheduled at time $t+m$, then $D_l(t+m) = 0$. Considering all the above cases for link $l$, it is easy to see that $\bar{\mathbb{E}}_{\mathbf{X}_t}[X_l(t)(D_l(t+m)-(1-\rho)D'_l(t+m))|\mathbf{Y}_t] \leq \bar{\mathbb{E}}_{\mathbf{X}_t}[X_l(t)(r+\rho(1-r))D_l(t+m)|\mathbf{Y}_t]$.

Since the above discussion holds for all links, we have that
$$\bar{\mathbb{E}}_{\mathbf{X}_t}[\mathbf{X}_t(\mathbf{D}_{t+m} - (1-\rho)\mathbf{D}'_{t+m})|\mathbf{Y}_t]$$
$$\leq (r + (1-r)\rho)\bar{\mathbb{E}}_{\mathbf{X}_t}[\mathbf{X}_t \mathbf{D}_{t+m}|\mathbf{Y}_t],$$
which implies that $\Psi_{\mathbf{Y}_t}^K \leq (r + (1-r)\rho)\Phi_{\mathbf{Y}_t}^K$. Using this inequality, the definition of $\theta$, and the fact that $\max(a - bx, x) \geq \frac{a}{1+b}$, we can show that
$$\theta \geq \frac{1 - \zeta' - \frac{\sqrt{N}\rho\alpha}{\delta\nu}}{1 + \frac{1-\delta}{\delta}(r + (1-r)\rho)}. \tag{29}$$

The term $\frac{\sqrt{N}\rho\alpha}{\delta\nu}$ in the right hand side of (29) can be made arbitrarily small by choosing the policy parameter $\alpha$ sufficiently small, which, as we show later, comes at the price of increasing the delay-bound proportionally to $\frac{1}{\alpha}$. Summarizing the preceding discussions, we have the following corollary.

**Corollary 3.** *Suppose the state of each link is a two-state Markov chain with transition probability $r$ and independent of the states of other links in the network. In addition, suppose the interference can be modeled by the node exclusive interference model or, more generally, by the $\kappa$-hop interference model. Finally, suppose the algorithm $A$ always returns a matching (or a $\kappa$-valid matching) with respect to the network graph $G$.*

*Then, for any input rate strictly inside $\theta_{min}\Gamma$, where*
$$\theta_{min} = \frac{1 - \zeta'}{1 + \frac{1-\delta}{\delta}((1-r)\rho + r)}.$$
*there exists a sufficiently small $\alpha$ such that the network is stabilized under LM-RSP. In other words, the region $\theta\Gamma$ contains the interior of the region $\theta_{min}\Gamma$ in the limit of small $\alpha$'s. In addition, when $\rho \ll \delta$, we have*
$$\theta_{lim} \geq \theta_{min} \simeq \frac{1 - \zeta}{1 + \frac{1-\delta}{\delta}r}.$$

The corollary essentially states that a *fixed fraction* of the capacity region $\Gamma$, regardless of the number of data flows $N$, can be stabilized by LM-RSP given that the pair $(\zeta, \delta)$ and the rate $r$ are fixed. Furthermore, it remarkably states that the *total channel variation rate* $1 - (1 - r)^N$, which is close to $Nr$ for small $r$'s, does not appear in the lower-bound fraction $\theta_{min}$, and what appears is the individual link variation rate $r$. As the last observation, note that the more restrictive is the interference model, i.e., when $\kappa$ becomes large in the $\kappa$-hop interference model, the smaller is the region $\Gamma$. However, the corollary assures that for a given $(\zeta, \delta)$ and $r$, the lower-bound fraction $\theta_{min}$ is not affected by the choice of $\kappa$, and thus, a fixed fraction of $\Gamma$ can be stabilized no matter how restrictive is the interference.

## VI. LM-RSP Delay Performance

In this section, we study the delay performance of LM-RSP. We start by introducing a few important parameters.

### A. Convergence Parameter $K$ and Norm Lower-Bound $B$

Here, we introduce two key parameters that play a central role in the delay analysis. The first is $K_\epsilon$ that essentially is a function of how fast channel states converge to their steady states, where the variable $\epsilon$ is used to measure the closeness of the input rate to the boundary of the region $\theta\Gamma$. In our analysis, $K_\epsilon$ determines the number of steps used in the Lyapunov drift-analysis. To formally define $K_\epsilon$, suppose a positive $\epsilon$ is given, and let $\epsilon_1 = \frac{1}{6}\frac{1}{D_{max}}\frac{\nu}{N}\frac{\epsilon}{4}$, $\epsilon_2 = \frac{\delta}{2}\epsilon_1$, and $\epsilon_3 = \frac{\epsilon}{4}$. We define $K_\epsilon$ as
$$K_\epsilon = 2\max(K_{1,\epsilon_1}, K_{2,\epsilon_2}^{(\gamma)}, D_{max}\frac{6}{\delta}\frac{N}{\nu}\frac{4}{\epsilon}, K_{\epsilon_3}^{(\theta)}), \tag{30}$$
where $K_{1,\epsilon_1}$ and $K_{2,\epsilon_2}^{(\gamma)}$ are defined in Section III-B, and $K_{\epsilon_3}^{(\theta)}$ is defined in Fact 3.

The second parameter is $B_\epsilon^K$, which acts as a lower-bound for the norm of $\|\mathbf{X}_t\|$, above which the Lyapunov drift becomes negative in our analysis. More specifically, if $\|\mathbf{X}_t\| \geq B_\epsilon^K$, then within the $K$ timeslots after time $t$, the inequalities in Property 2 and Fact 2 hold with high probability. To formally define $B_\epsilon^K$, suppose for a given $K$ and a positive $\epsilon$, $\epsilon_4$ and $\epsilon_5$ are defined by $\epsilon_4 = \frac{1}{6}\frac{1}{D_{max}}\frac{\nu}{N}\frac{\epsilon}{4}\frac{\delta}{2K}$ and $\epsilon_5 = \frac{\epsilon_4}{4}$. Let $\bar{A}_{\epsilon_4} = A_{\epsilon_4} + \sqrt{N}D_{max}$, where $A_{\epsilon_4}$ is defined by Fact 1. We define $B_\epsilon^K$ as
$$B_\epsilon^K = \max(B_{1,\epsilon_5}^C, B_{2,\epsilon_5}^C), \tag{31}$$
where $C = K\bar{A}_{\epsilon_4}$, and $B_{1,\epsilon_5}^C$ and $B_{2,\epsilon_5}^C$ are defined by Property 2 and Fact 2, respectively.

## B. Big $\mathcal{O}$ Notation

As a notational convenience in our following analysis and discussions, we use the big $\mathcal{O}$ notation with multiple variables. In such cases, we assume the ordinary big $O$ notation holds individually for each present *independent* variable as it takes its limiting value. In particular, we consider the scaling behaviors, as $N \to \infty$, $\rho \to 0$, $\alpha \to 0$, $\delta \to 0$, $\zeta \to 1$, or $\epsilon \to 0$, where $\epsilon$ is defined in Theorem 1.

## C. Theorem on Average Expected Queue-Lengths

The following is the main theorem on the average expected queue-lengths.

**Theorem 2.** *Under the assumptions in Theorem 1, the expected queue-lengths satisfy the following:*

$$\overline{\sum_{1 \leq l \leq N} X_l} = \limsup_{T \to \infty} \frac{1}{T+1} \sum_{t=0}^{T} \Big[ \sum_{l=1}^{N} \mathbb{E}\big[X_l(t)\big] \Big]$$
$$\leq B(\sqrt{N} + 2\frac{N}{\nu\epsilon}\|\mathbf{a}\|) + \mathcal{O}\Big(\frac{KN^2}{\delta\epsilon}\Big),$$

*where $\nu$ is defined in (19), $\epsilon$ is given by Theorem 1, $\delta$ is defined in Property 1, $K = K_\epsilon$, and $B = B_\epsilon^K$.*

*Proof.* The proof of the theorem is given in the appendix. $\square$

To gain insights into the delay performance of LM-RSP using the above theorem, we need to study the properties of the parameters $K_\epsilon$ and $B_\epsilon^K$, where $\epsilon$ is defined in Theorem 1. Note that this value of $\epsilon$ is used to determine $\epsilon_i$'s, $1 \leq i \leq 5$, defining $K_\epsilon$ and $B_\epsilon^K$. We start by considering the definition of $B_\epsilon^K$ given in (31), implying that $B_\epsilon^K \geq B_{2,\epsilon_5}^C$. It is easy to see that since $f(\cdot)$ is linear in the range $[-\rho, \rho]$, we can set $B_{2,\epsilon_5}^C = C(\frac{8\sqrt{N}D_{max}}{2\rho\alpha\epsilon_5} + 1)$. It is also easy to verify that this form for $B_{2,\epsilon_5}^C$ is indeed necessary when constant multiplicative factors are ignored. Hence, by the definitions for $\epsilon_4$ and $\epsilon_5$, and that $C = K\bar{A}_{\epsilon_4}$, all given in Section VI-A, we have

$$B_{2,\epsilon_5}^C = \Omega(\frac{N^{1.5}K^2}{\rho\alpha\delta\epsilon}). \tag{32}$$

By (30), we have $K > \frac{D_{max}N}{\delta\nu\epsilon}$, and by (23), we have $\rho\alpha < \frac{\delta\nu}{\sqrt{N}}$. Using these inequalities, the equality (32), and that $B_\epsilon^K \geq B_{2,\epsilon_5}^C$, it is easy to see $B(\sqrt{N} + 2\frac{N}{\nu\epsilon}\|\mathbf{a}\|)$ dominates the term $\mathcal{O}\Big(\frac{KN^2}{\delta\epsilon}\Big)$. We, therefore, have the following corollary.

**Corollary 4.** *The dominant term in the average queue-length is $B(\sqrt{N} + 2\frac{N}{\nu\epsilon}\|\mathbf{a}\|)$.*

Note that the dominant term implicitly depends on $K = K_\epsilon$ through the term $B = B_\epsilon^K$. Therefore, to have a specific bound on the average queue-lengths, we need to also study $K_\epsilon$. By definition, $K_\epsilon$ depends on $K_{1,\epsilon_1}$, $K_{2,\epsilon_2}^{(\gamma)}$, and $K_{\epsilon_3}^{(\theta)}$, which can be considered as convergence rates. The first two rates are essentially the convergence rate of the channel process to its steady state. The third rate $K_{\epsilon_3}^{(\theta)}$ depends on both the channel convergence rate and the update policy. Hence, towards having a specific average delay-bound, we need to focus on a particular channel model, as discussed next.

## D. Specific Delay Bound for Markovian Channels

As a special example, suppose the channel state is a Markov chain as described in Section V-C.2, where the state of each link is a two-state Markov chain with transition probability $r$ and independent of states of other links in the network. The following is the key lemma [31] that we use to study the convergence rates of Markov chains.

**Lemma 1.** *Suppose a Markov chain is defined on the finite state space $\mathcal{X}$ with transition probabilities $P(x,y)$, where $x \in \mathcal{X}$ and $y \in \mathcal{X}$. Let $\pi_k$ be the distribution after $k$ transitions given an initial distribution $\pi_0$. Then, given any initial distribution $\pi_0$ and stationary distribution $\pi$, we have*

$$\sum_{x \in \mathcal{X}} |\pi_k(x) - \pi(x)| \leq 2(1-\beta)^{\lfloor \frac{k}{k_0} \rfloor},$$

*where $k_0$ is a positive integer and*

$$\beta = \sum_{y \in \mathcal{X}} \min_{x \in \mathcal{X}} P^{k_0}(x,y),$$

*where $P^{k_0}$ denotes the transition probability after $k_0$ transitions.*

We first concentrate on $K_{\epsilon_3}^{(\theta)}$. Since the channel is Markovian, by the definition of $K_{\epsilon_3}^{(\theta)}$ and the discussion leading to (24) and (25), we see that $K_{\epsilon_3}^{(\theta)}$ depends on how fast $\Psi_{\mathbf{Y}_t}^K$ and $\Phi_{\mathbf{Y}_t}^K$ converge to $\Psi_{\widetilde{\mathbf{Y}}_t}^\infty$ and $\Phi_{\widetilde{\mathbf{Y}}_t}^\infty$, respectively. Recall that in $\Psi_{\mathbf{Y}_t}^K$ and $\Phi_{\mathbf{Y}_t}^K$ queue dynamics are ignored, and $\mathbf{I}(t+k)$, $k > 0$, is updated by setting $\mathbf{X}(t+k) = \mathbf{X}(t)$. Using this along with the discussion leading to (24), we see that the process $\{(\mathbf{s}_{t+k}, \mathbf{I}_{t+k}), k \geq 0\}$ is also a Markov chain on the space $\mathcal{S} \times \mathcal{I}$. Our goal is to find an appropriate value for $\beta$ to apply Lemma 1 to this Markov chain.

Consider the $l_{\text{th}}$ link whose state $s_l$ is by itself a two-state Markov chain with steady-state probabilities $\pi(b) = \pi(g) = 0.5$. Setting $k_0 = 1$, we can use Lemma 1 to show that

$$\sum_{s \in \{b,g\}} |\pi_k(s_l = s) - \pi(s_l = s)| < 2(1-\beta_l)^k, \tag{33}$$

where $\beta_l = 2r$ if $r < 0.5$, and $\beta_l = 2 - 2r$ otherwise. Since there are $N$ links with independent states, the above inequality indicates that after $k$ transitions after time $t$, with probability at least $(\frac{1}{2} - (1-\beta_l)^k)^N$, the state $\mathbf{s}_{t+k}$ satisfies $\mathbf{s}_{t+k} = \mathbf{s}$, where $\mathbf{s}$ can be any state in $\mathcal{S}$. On the other hand, by Property 1, with probability at least $\delta$, inequality (11) with $\mathbf{X} = \mathbf{X}_t$ and $\mathbf{s} = \mathbf{s}_t$ holds for any time-slot $t$. This implies that for any state $\mathbf{s} \in \mathcal{S}$, there exists a set $A_{\mathbf{X}_t, \mathbf{s}}$, $A_{\mathbf{X}_t, \mathbf{s}} \subset \mathcal{I}$, such that for all $\mathbf{I} \in A_{\mathbf{X}_t, \mathbf{s}}$, inequality (11) holds for $\mathbf{I}' = \mathbf{I}$ and $\mathbf{X} = \mathbf{X}_t$, and

$$\sum_{\mathbf{I} \in A_{\mathbf{X}_t, \mathbf{s}}} \mu_{\mathbf{X}_t, \mathbf{s}}(\mathbf{I}' = \mathbf{I}) \geq \delta, \tag{34}$$

where $\mu_{\mathbf{X}_t, \mathbf{s}}$ is defined in Section IV-A.

Now, suppose at time $t+k$, for a given $k$, the algorithm $A$ chooses the schedule $\mathbf{I}$ that belongs to the set $A_{\mathbf{X}_t, \mathbf{s}_{t+k}}$[8]. This happens with probability $\mu_{\mathbf{X}_t, \mathbf{s}_{t+k}}(\mathbf{I}' = \mathbf{I})$. In this case, we have $\varphi(t+k) \geq -\zeta$. Assuming $\rho > 2\zeta$, by the update rule, we have that with probability at least $\frac{1}{4}$, $\mathbf{I}_{t+k} = \mathbf{I}'(t+k) = \mathbf{I}$. This and the discussion in the previous paragraph imply that for any channel state $\mathbf{s}$ and $\mathbf{I} \in A_{\mathbf{X}_t, \mathbf{s}}$, *regardless* of the initial state $(\mathbf{s}_t, \mathbf{I}_t)$, after $k$ time-slots, with probability at least $P_{min}^k((\mathbf{s}, \mathbf{I})) = \frac{1}{4}(\frac{1}{2} - (1 - \beta_l)^k)^N \mu_{\mathbf{X}_t, \mathbf{s}}(\mathbf{I}^r = \mathbf{I})$, the chain will be at state $(\mathbf{s}_{t+k}, \mathbf{I}_{t+k}) = (\mathbf{s}, \mathbf{I})$. Using this lower bound as the minimum transition probability in the expression for $\beta$ and replacing $k$ with $k_0$, we can show that $\beta$ for the Markov chain $\{(\mathbf{s}_{t+i}, \mathbf{I}_{t+i}), i \geq 0\}$ satisfies

$$
\begin{aligned}
\beta &\geq \sum_{(\mathbf{s}, \mathbf{I}): \mathbf{s} \in \mathcal{S}, \mathbf{I} \in A_{\mathbf{X}_t, \mathbf{s}}} P_{min}^{k_0}((\mathbf{s}, \mathbf{I})) \\
&= \sum_{\mathbf{s} \in \mathcal{S}} \sum_{\mathbf{I} \in A_{\mathbf{X}_t, \mathbf{s}}} (\frac{1}{2} - (1 - \beta_l)^{k_0})^N \frac{1}{4} \mu_{\mathbf{X}_t, \mathbf{s}}(\mathbf{I}^r = \mathbf{I}) \\
&= \frac{1}{4} \sum_{\mathbf{s} \in \mathcal{S}} (\frac{1}{2} - (1 - \beta_l)^{k_0})^N \sum_{\mathbf{I} \in A_{\mathbf{X}_t, \mathbf{s}}} \mu_{\mathbf{X}_t, \mathbf{s}}(\mathbf{I}^r = \mathbf{I}) \\
&\geq \frac{1}{4} \delta (1 - 2(1 - \beta_l)^{k_0})^N,
\end{aligned}
\tag{35}
$$

where, to obtain the last inequality, we have used (34) and that $|\mathcal{S}| = 2^N$. Hence, we have obtained a lower-bound for $\beta$ given any $k_0$. Next, we use this bound along with Lemma 1 to study the convergence of $\Phi_{\mathbf{Y}_t}^K$ to $\Phi_{\mathbf{Y}_t}^\infty$.

---

[8] In the context of this discussion, since we are focusing on $\Phi_{\mathbf{Y}_t}^K$, by its definition, queue dynamics after time $t$ are ignored, and thus $A_{\mathbf{X}_{t+k}, \mathbf{s}_{t+k}} = A_{\mathbf{X}_t, \mathbf{s}_{t+k}}$.

Suppose $\pi_k(\cdot, \cdot)$ is the distribution of the Markov chain $\{(\mathbf{s}_{t+i}, \mathbf{I}_{t+i}), i \geq 0\}$ after $k$ transitions, given $\mathbf{s}_t$ and $\mathbf{I}_t$. From the definition of $\Phi_{\mathbf{Y}_t}^K$ and $\Phi_{\mathbf{Y}_t}^\infty$, given in (18) and (25), respectively, we have that

$$
\begin{aligned}
|\Phi_{\mathbf{Y}_t}^K - \Phi_{\mathbf{Y}_t}^\infty| &= \frac{1}{K \Upsilon(\mathbf{X}_t)} \\
&\qquad \Big| \sum_{k=0}^{K-1} \sum_{\mathbf{s}, \mathbf{I}} (\pi_k(\mathbf{s}, \mathbf{I}) - \pi(\mathbf{s}, \mathbf{I})) \mathbf{X}_t \mathbf{D}(\mathbf{s}, \mathbf{I}) \Big| \\
&< \frac{\sqrt{N}}{K \nu \|\mathbf{X}_t\|} \sum_{k=0}^{K-1} \sum_{\mathbf{s}, \mathbf{I}} \Big| (\pi_k(\mathbf{s}, \mathbf{I}) - \pi(\mathbf{s}, \mathbf{I})) \Big| \sqrt{N} \|\mathbf{X}_t\| D_{max} \\
&\leq \frac{N D_{max}}{K \nu} \sum_{i=0}^{\lceil \frac{k_0}{K} \rceil} k_0 2(1 - \beta)^i \leq \frac{2 N D_{max}}{\nu} \frac{k_0}{K} \frac{1}{\beta},
\end{aligned}
\tag{36}
$$

where the first inequality follows from (2) and (20), and the second inequality is a direct result of Lemma 1. Let

$$
k_0 = \begin{cases} \lceil \frac{\ln(4N)}{-\ln(1 - \beta_l)} \rceil, & r \neq 0.5 \\ 1, & r = 0.5 \end{cases},
$$

which implies that $(1 - \beta_l)^{k_0} \leq \frac{1}{4N}$. This inequality and (35) further imply that $\beta \geq \frac{\delta}{8}$. Therefore, from (36), we have that

$$
|\Phi_{\mathbf{Y}_t}^K - \Phi_{\mathbf{Y}_t}^\infty| < \frac{16 N D_{max}}{\nu} \frac{k_0}{K \delta}.
\tag{37}
$$

We can obtain a similar upper-bound for $|\Psi_{\mathbf{Y}_t}^K - \Psi_{\mathbf{Y}_t}^\infty|$ by considering separately two Markov chains corresponding to the pairs $(\mathbf{s}_{t+k}, \mathbf{I}_{t+k})$ and $(\mathbf{s}_{t+k+1}, \mathbf{I}_{t+k})$, $k \geq 0$, respectively. Specifically, we can show that

$$
|\Psi_{\mathbf{Y}_t}^K - \Psi_{\mathbf{Y}_t}^\infty| < \frac{1}{K} \Big( 1 + \frac{32 N D_{max}}{\nu} \frac{k_0}{\delta} \Big).
$$

Using this inequality, the one in (37), and the definition of $\theta$, it is easy to see that we can set

$$
K_{\epsilon_3}^{(\theta)} = \Big\lceil \frac{1}{\delta \epsilon_3} \Big(1 + \frac{32 N D_{max}}{\nu} \frac{k_0}{\delta}\Big) \Big\rceil = \Theta\Big(\frac{N k_0}{\epsilon_3 \delta^2}\Big).
\tag{38}
$$

Using similar approaches, we can show that $K_{1,\epsilon}$ and $K_{2,\epsilon_2}^{(\gamma)}$ can be chosen as

$$
K_{1,\epsilon_1} = \Big\lceil \frac{4 k_0}{\epsilon_1} \Big\rceil = \Theta\Big(\frac{k_0}{\epsilon_1}\Big)
\tag{39}
$$

and

$$
K_{2,\epsilon_2}^{(\gamma)} = \Big\lceil \frac{8 k_0}{\epsilon_2} + \frac{\ln(2)}{\ln((1 - \delta)^{-1})} \Big\rceil = \Theta\Big(\frac{k_0}{\epsilon_2}\Big).
\tag{40}
$$

Finally, using the definition of $K_\epsilon$ and equalities (38), (39), and (40), we can easily verify that we can choose $K_\epsilon$ such that

$$
K_\epsilon = \Theta\Big(\frac{N k_0}{\delta^2 \epsilon}\Big).
\tag{41}
$$

After studying $K_\epsilon$ for a Markovian channel model, we now return to finding an upper-bound for the dominant term given by Corollary 4. Suppose arrivals are limited by a constant $\bar{A}_{max}$[9], i.e., $A_l(t) < \bar{A}_{max}$, $1 \leq l \leq N$. Using this assumption, we can set $\bar{A}_{\epsilon_4} = \sqrt{N}(A_{max} + D_{max})$, where $\bar{A}_{\epsilon_4}$ is given in the definition for $B_\epsilon^K$. Using the same discussion leading to (32) while not excluding the effect of $\bar{A}_{\epsilon_4}$, we see that $\bar{A}_{\epsilon_4}$ contributes a $\sqrt{N}$ term into $B_{2,\epsilon_5}^C$, and therefore, we can have

$$
B_{2,\epsilon_5}^C = \Theta\Big(\frac{N^2 K^2}{\rho \alpha \delta \epsilon}\Big).
\tag{42}
$$

In addition, suppose

$$
B_{1,\epsilon_5}^C = \mathcal{O}\Big(\frac{N^2 K^2}{\rho \alpha \delta \epsilon}\Big).
\tag{43}
$$

This for instance is the case where the algorithm $A$ chooses candidate schedules from a fixed set with equal probabilities. In this particular case, the distribution of $\mathbf{I}^r$ does not depend on $\mathbf{X}_t$, and thus, $B_{1,\epsilon_5}^C$ can be assumed to be any positive real number. Note that $B_{1,\epsilon_5}^C$ depends on a specific implementation of the algorithm $A$, a topic that is not the focus of this paper.

Recall that $B = B_\epsilon$ and $K = K_\epsilon$. It follows from (31), (41), (42), and (43) that

$$
B(\sqrt{N} + 2\frac{N}{\nu \epsilon} \|\mathbf{a}\|) = \mathcal{O}\Big(\frac{N^{4.5} k_0^2}{\rho \alpha \delta^5 \epsilon^3}\Big) + \mathcal{O}\Big(\frac{N^5 k_0^2 \|\mathbf{a}\|}{\rho \alpha \delta^5 \epsilon^4}\Big).
\tag{44}
$$

In addition, since $-\ln(1 - \beta_l) \geq \beta_l$, we have $k_0 \leq \frac{\ln(4N)}{\beta_l}$, which along with equality (44) and Corollary 4 leads to the following corollary.

**Corollary 5.** *Suppose the state of each link is a two-state Markov chain with transition probability $r$ and independent of the states of other links in the network. In addition, suppose arrivals are limited by a suitably large constant, $B_{1,\epsilon_5}^C$ satisfies $B_{1,\epsilon_5}^C = \mathcal{O}(\frac{N^2 K^2}{\rho \alpha \delta \epsilon})$, and $\rho > 2\zeta$. Then, assuming $r \leq 0.5$, the average queue-lengths satisfy the following:*

$$
\begin{aligned}
\overline{\sum_{1 \leq l \leq N} X_l} &= \mathcal{O}\Big(\frac{N^{4.5}(\ln(N))^2}{\rho \alpha \delta^5 \epsilon^3 r^2}\Big) \\
&\quad + \mathcal{O}\Big(\frac{N^5 (\ln(N))^2 \|\mathbf{a}\|}{\rho \alpha \delta^5 \epsilon^4 r^2}\Big).
\end{aligned}
$$

*If $r > 0.5$, the same result holds except that the terms $r^2$ should be replaced with $(1 - r)^2$.*

We are finally at a stage to study how delay scales according to various network- or policy-related parameters. Suppose the input rate is $\mathbf{a} = (\lambda_1, \cdots, \lambda_N)$ strictly inside $\theta\Gamma$, and consider the coefficients $\beta_{\mathbf{s}, \mathbf{I}}$'s corresponding to the rate $\mathbf{a}$, as specified in Theorem 1. Let $\theta_{\mathbf{a}} = \max_{\mathbf{s} \in \mathcal{S}} \sum_{\mathbf{I} \in \mathcal{I}} \beta_{\mathbf{s}, \mathbf{I}}$, where by the theorem we have $\theta_{\mathbf{a}} < \theta$. Based on the definition of $\theta$ and $\theta_{\mathbf{a}}$, it is clear that the rate vector $\frac{\theta}{\theta_{\mathbf{a}}} \mathbf{a}$ belongs to the boundary of $\theta\Gamma$. Since the region $\theta\Gamma$ serves as the *reference* stability region, the rate $\mu_l = \frac{\theta}{\theta_{\mathbf{a}}} \lambda_l$, $1 \leq l \leq N$, which is the $l$th element of the vector $\frac{\theta}{\theta_{\mathbf{a}}} \mathbf{a}$, can be regarded as the *effective service rate* for the $l$th link. This provides the motivation to define $\varsigma$ as $\varsigma = \frac{\theta_{\mathbf{a}}}{\theta}$ and to consider $\varsigma$ as the effective load for each link.

---

[9]Recall that we earlier in Section III-A introduced $\bar{A}_{max}$ as the upper-bound for the second moments of the arrival process.

Based on this definition, we have $\lambda_l = \varsigma\mu_l$, $1 \le l \le N$, and for $\epsilon$, as defined in Theorem 1 and used throughout previous discussions, we have that $\epsilon = (1-\varsigma)\theta$. From Little's theorem, we have that the overall average delay for each packet, denoted by $\bar{D}$, is given by $1/(\sum_{l=1}^{N} \lambda_l)\overline{\sum_{1 \le l \le N} X_l}$, which along with the inequality $\sum_{l=1}^{N} \lambda_l \ge \|\mathbf{a}\|$ and Corollary 5 leads to the following corollary.

**Corollary 6.** *Suppose the state of each link is a two-state Markov chain with transition probability $r$ and independent of the states of other links in the network. In addition, suppose arrivals are limited by a suitably large constant, $B_{1,\epsilon_5}^C$ satisfies $B_{1,\epsilon_5}^C = \mathcal{O}(\frac{N^2 K^2}{\rho\alpha\delta\epsilon})$, and $\rho > 2\varsigma$. Then, assuming $r \le 0.5$, the overall average queue delay $\bar{D}$ satisfies*

$$\bar{D} \triangleq \frac{\overline{\sum_{1 \le l \le N} X_l}}{\sum_{l=1}^{N} \lambda_l} = \mathcal{O}\left(\frac{N^{4.5}(\ln(N))^2}{\rho\alpha\delta^5(1-\varsigma)^3\theta^3 r^2 \varsigma\mu_{(t)}}\right)$$
$$+ \mathcal{O}\left(\frac{N^5(\ln(N))^2}{\rho\alpha\delta^5(1-\varsigma)^4\theta^4 r^2}\right),$$

*where $\mu_{(t)}$ is the total service rate and is given by $\mu_{(t)} = \sum_{1 \le l \le N} \mu_l$. If $r > 0.5$, the same result holds except that the terms $r^2$ should be replaced with $(1-r)^2$.*

Note that the format of the obtained delay-bound is similar to the average delay for the M/M/1 queue, which is $\frac{1}{(1-\varsigma)\mu}$, with $\varsigma$ as the load and $\mu$ as the service rate. Remarkably, the corollary states that delay is polynomially bounded as the variables of interest, including the number of data flows $N$, take their limiting values. In particular, we see that delay is $\mathcal{O}(\frac{1}{r^2})$ as the link variation rate $r$ takes smaller values, and is $\mathcal{O}(\frac{1}{\delta^5})$ as $\delta \to 0$. In the next section, we consider both the throughput and delay performance of LM-RSP.

## VII. Joint Throughput-Delay Performance

In this section, with the help of the corollaries provided earlier, we investigate the throughput and delay scaling as the variables of interest take their limiting values. As discussed earlier, Theorem 1 and Corollary 1 state that LM-RSP can stabilize a fraction of the capacity region $\Gamma$. Corollary 3 further shows that the policy can stabilize a *fixed* fraction $\theta_{min}$ of the capacity region regardless of the number data flows $N$ if the rest of parameters are fixed. However, as expected and inherently present in Theorem 2, the delay-bound increases with $N$. Specifically, Corollary 4 characterizes the dominant term in the delay-bound, leading to Corollary 6 that states delay is bounded by a polynomially increasing function of $N$.

An interesting trade-off occurs when parameters $\rho$ and $\alpha$ take vanishingly small values. Recall that these parameters must be positive for continuity purposes. From Corollaries 1-3, we observe that as $\alpha$ and $\rho$ take smaller values, the stability region $\theta\Gamma$ is ensured to contain a region arbitrarily close to the region $\theta_{lim}\Gamma$ or, for Markovian channels, the region $\theta_{min}\Gamma$. Corollary 6, on the other hand, shows that this comes at the price of increasing the delay-bound proportionally to $\frac{1}{\alpha\rho}$.

A similar trade-off exists when the channel states become increasingly correlated. Specifically, as discussed in Section V-C.1, increasing channel correlation increases $\theta$ and, thus, expands the stability region. We also discussed that channel correlation helps the policy compensate for a small $\delta$. This is more explicitly stated in Corollary 3, which shows that for a particular Markovian channel process, as the link variation rate $r$ decreases, the lower-bound region $\theta_{min}\Gamma$ expands almost proportionally to $(1 + (r + \rho)\frac{1-\delta}{\delta})^{-1}$. In particular, in the limit $r \to 0$ and $\rho \to 0$, for $\zeta = 0$ and any positive $\delta$, we have $\theta_{min} \to 1$, and thus, throughput-optimality can be achieved, similar to the observation in [5]. However, as shown in Corollary 6, this makes the delay-bound increase proportionally to $r^{-2}$ as $r \to 0$.

Finally, we focus on the pair $(\zeta, \delta)$. Part (b) of Theorem 1 and also Corollaries 1-3 all state that $\zeta$ can directly affect $\theta$ through the term $(1 - \zeta') = (1 - \rho)(1 - \zeta)$ or, otherwise, the term $(1 - \zeta)$. By Corollary 6, since for a given load factor $\varsigma$, the delay-bound increases proportionally to $\frac{1}{\theta^4}$ as $\theta \to 0$, we have that the larger is $\zeta$, the smaller is $\theta$, and the larger should be the delay. Therefore, increasing $\zeta$ has negative effects on both throughput and delay. Recall that the parameter $\delta$ is the least probability that the candidate schedules are within $\zeta$-neighborhood of the optimal schedules. Therefore, as is clear in the definition for $\theta$, smaller values for $\delta$ decrease $\theta$. Specifically, Corollary 3 shows that $\theta_{min}$, as a lower-bound for $\theta$, decreases almost proportionally to $\frac{\delta}{\delta+\rho+r}$ as $\delta$ approaches zero. Decreasing $\delta$ has also an adverse effect on the delay-bound since by Corollary 6, the bound can increase proportionally to $\frac{1}{\delta^5}$ as $\delta$ decreases. As the final remark, note that if it is possible to increase $\delta$, at the expense of increasing the complexity of algorithm $A$, it can be sufficient to make sure that $\delta$ has the same order as $r$. This is an intuitive observation and a result of Corollary 3, which states that for Markovian channels, we may have $\theta_{min} \simeq \frac{\delta}{\delta+(1-\delta)r}(1-\zeta)$. For instance, when $\delta = r$, if the update rule is not used, the scaling of the capacity region is $r(1-\zeta)$ whereas using LM-RSP can ensure $\theta_{min} \ge \frac{1}{2}(1-\zeta)$, which implies a significant throughput improvement especially when $r \ll 0.5$.

## VIII. Conclusion

In this paper, we have studied the stability region and delay performance of a linear-memory randomized scheduling policy LM-RSP for networks with time-varying channels. LM-RSP uses an update rule along with a randomized algorithm that with probability at least $\delta$ finds a candidate schedule vector that is within $\zeta$-neighborhood of optimality. The complexity of LM-RSP depends on the complexity of the randomized algorithm and, in particular, may be linear. We have proved that LM-RSP can stabilize a scaled version (fraction) of the capacity region and quantified the corresponding scaling factor as a function of the parameters in LM-RSP and the limiting behavior of rate changes due to channel variations. Furthermore, we have provided an average delay-bound for general ergodic channel processes. For a particular class of Markovian channels, we have shown that the average delay is $\mathcal{O}(\frac{1}{r^2})$, as $r \to 0$, where $r$ is the link (individual channel) variation rate, and is bounded by a polynomially increasing function of the number of data flows. In addition, for this class of channels, we have shown that a minimum fraction $\frac{\delta}{\delta+r}(1-\zeta)$ of the capacity region can be stabilized. Our results also indicate that while the minimum fraction decreases linearly as $\delta \to 0$, the delay may increase as $\frac{1}{\delta^5}$, and therefore, the effect of $\delta$ on delay may be more severe than the one on the stability region. The results in this paper are promising and motivate future research as they indicate that even when channels are time-varying, using randomized policies can help stabilize a *predictable* fraction of the capacity region, in networks with limited computation power and memory resource, while assuring a polynomially-bounded delay.

## Appendix
## Proofs of Theorem 1 and Theorem 2

The proofs use several lemmas that are provided in a separate appendix at the end of the paper.

*Proof of part (a) of Theorem 1.* Since $\mathbf{a}$ is strictly inside $\theta\Gamma$, there should be a rate $\mathbf{a}_1$ inside $\Gamma$ such that $\mathbf{a} = \theta\mathbf{a}_1$. By the definition of $\Gamma$, it is easy to see that there should exist non-negative constants $\beta'_{\mathbf{s},\mathbf{I}}$'s such that for all $\mathbf{s} \in \mathcal{S}$, $\sum_{\mathbf{I} \in \mathcal{I}} \beta'_{\mathbf{s},\mathbf{I}} < 1$ and $\mathbf{a}_1 = \sum_{\mathbf{s} \in \mathcal{S}} \pi(\mathbf{s}) \sum_{\mathbf{I} \in \mathcal{I}} \beta'_{\mathbf{s},\mathbf{I}} \mathbf{D}(\mathbf{s}, \mathbf{I})$. Using this equality and setting $\beta_{\mathbf{s},\mathbf{I}} = \theta\beta'_{\mathbf{s},\mathbf{I}}$, we see that for these choices of $\beta_{\mathbf{s},\mathbf{I}}$'s, $\mathbf{a}$ satisfies (27), and $\epsilon$ as defined by $\epsilon = \theta - \max_{\mathbf{s} \in \mathcal{S}} \sum_{\mathbf{I} \in \mathcal{I}} \beta_{\mathbf{s},\mathbf{I}}$ is positive, as required. □

*Proof of part (b) of Theorem 1.* We use a K-step drift analysis to prove part (b) of the theorem. The main difficulty here is to properly use the properties of LM-RSP in the drift analysis. Consider the following Lyapunov function:

$$V(\mathbf{Y}_t) = \sum_{l=1}^{N} X_l(t)^2.$$

We can write a K-step drift as follows:

$$\Delta(K)_t = \mathbb{E}[V(\mathbf{Y}_{t+K}) - V(\mathbf{Y}_t)|\mathbf{Y}_t]$$
$$= \sum_{k=0}^{K-1} \mathbb{E}[V(\mathbf{Y}_{t+k+1}) - V(\mathbf{Y}_{t+k})|\mathbf{Y}_t]$$
$$= \sum_{k=0}^{K-1} \mathbb{E}[(\mathbf{X}_{t+k+1} + \mathbf{X}_{t+k})(\mathbf{X}_{t+k+1} - \mathbf{X}_{t+k})|\mathbf{Y}_t]$$
$$= \sum_{k=0}^{K-1} \mathbb{E}[2\underbrace{\mathbf{X}_{t+k}(\mathbf{A}_{t+k} - \mathbf{D}_{t+k})}_{\delta_{1,k}} + 2\underbrace{\mathbf{X}_{t+k}\mathbf{U}_{t+k}}_{\delta_{2,k}} +$$
$$\underbrace{(\mathbf{A}_{t+k} - (\mathbf{D}_{t+k} - \mathbf{U}_{t+k}))(\mathbf{A}_{t+k} - (\mathbf{D}_{t+k} - \mathbf{U}_{t+k}))}_{\delta_{3,k}}|\mathbf{Y}_t]. \quad (45)$$

Based on the above expression, there are three main summations, corresponding to $\delta_{1,k}$, $\delta_{2,k}$, and $\delta_{3,k}$, respectively, each of which should be upper-bounded appropriately. To upper-bound the summation over $\delta_{2,k}$, note that by the definition of $\mathbf{U}(t)$, $U_i(t) \leq D_{max}$, and if $X_i(t) > D_{max}$, then $U_i(t) = 0$. Therefore,

$$\sum_{k=0}^{K-1} \mathbb{E}[\delta_{2,k}|\mathbf{Y}_t] \leq \sum_{k=0}^{K-1} ND_{max}^2 = KND_{max}^2. \quad (46)$$

As for the summation over $\delta_{3,k}$, we have that

$$\sum_{k=0}^{K-1} \mathbb{E}[\delta_{3,k}|\mathbf{Y}_t] \leq \sum_{k=0}^{K-1} \mathbb{E}[\mathbf{A}_{t+k}\mathbf{A}_{t+k}|\mathbf{Y}_t] + \mathbb{E}[\mathbf{D}_{t+k}\mathbf{D}_{t+k}|\mathbf{Y}_t]$$
$$\leq K\mathbb{E}[\|\mathbf{A}\|^2] + KND_{max}^2.$$

The heart of the proof, however, lies in deriving an upper-bound for the summation over $\delta_{1,k}$. It requires Lemmas 2-7, which are listed at the end of the proof of theorem. First, consider the following straightforward observations:

$$\mathbb{E}[\mathbf{X}_{t+k}\mathbf{A}_{t+k}|\mathbf{Y}_t] \leq \mathbb{E}[\mathbf{X}_t\mathbf{A}_{t+k}|\mathbf{Y}_t] + \mathbb{E}[\sum_{i=0}^{k-1} \mathbf{A}_{t+i}\mathbf{A}_{t+k}|\mathbf{Y}_t]$$
$$= \mathbf{X}_t\mathbf{a} + k\|\mathbf{a}\|^2, \quad (47)$$

and

$$\sum_{k=0}^{K-1} \mathbb{E}[\mathbf{X}_t\mathbf{A}_{t+k} - \mathbf{X}_{t+k}\mathbf{D}_{t+k}|\mathbf{Y}_t]$$
$$= K\mathbf{X}_t\mathbf{a} - \sum_{k=0}^{K-1} \mathbb{E}[\mathbf{X}_{t+k}\mathbf{D}_{t+k}^*|\mathbf{Y}_t]$$
$$+ \sum_{k=0}^{K-1} \mathbb{E}[\mathbf{X}_{t+k}(\mathbf{D}_{t+k}^* - \mathbf{D}_{t+k})|\mathbf{Y}_t]. \quad (48)$$

Since by the assumption in the theorem $\mathbf{a}$ is strictly inside $\theta\Gamma$, part (a) of the theorem holds. Specifically, there exist non-negative constants $\beta_{\mathbf{s},\mathbf{I}}$'s such that equality (27) holds, and $\epsilon$ as defined in part (a) is positive, i.e.,

$$\epsilon = \theta - \max_{\mathbf{s} \in \mathcal{S}} \sum_{\mathbf{I} \in \mathcal{I}} \beta_{\mathbf{s},\mathbf{I}} > 0. \quad (49)$$

To use the results of Lemmas 2-6, suppose[10] $\epsilon_1 = \epsilon_4 = \epsilon'_4 = \frac{1}{6}\frac{1}{D_{max}}\frac{\nu}{N}\frac{\epsilon}{4}$, $\epsilon_5 = \frac{1}{6}\frac{1}{D_{max}}\frac{\nu}{N}\frac{\epsilon}{4}\frac{\delta}{2}$, and $\epsilon_7 = \frac{\epsilon}{4}$, where $\epsilon$ is given by (49). We set

$$K = 2\max(K_{1,\epsilon_1}, K_{2,\epsilon_5}^{(\gamma)}, D_{max}\frac{6}{\delta}\frac{N}{\nu}\frac{4}{\epsilon}, K_{\epsilon_7}^{(\theta)}) = K_\epsilon, \quad (50)$$

where the last equality follows from the definition of $K_\epsilon$ given in (30). In addition, let $\epsilon_2 = \epsilon'_2 = \frac{1}{6}\frac{1}{D_{max}}\frac{\nu}{N}\frac{\epsilon}{4}\frac{\delta}{2R}$, $\epsilon_3 = \epsilon'_3 = \frac{1}{6}\frac{1}{D_{max}}\frac{\nu}{N}\frac{\epsilon}{4}\frac{\delta}{8R}$, and $B = B_{\epsilon_2,\epsilon_3}^K$, where $B_{\epsilon_2,\epsilon_3}^K$ is defined in Lemma 5. These choices imply that

$$B = B_\epsilon^K, \quad (51)$$

where $B_\epsilon^K$ is defined in (31).

With these choices, we can put together the results in (45), (46), (47), (48) to show that

$$\Delta(K)_t \leq 2KND_{max}^2 + K\mathbb{E}[\|\mathbf{A}\|^2] + KND_{max}^2$$
$$+ 2\sum_{k=0}^{K-1} \mathbb{E}[\mathbf{X}_{t+k}(\mathbf{A}_{t+k} - \mathbf{D}_{t+k})|\mathbf{Y}_t]$$
$$\leq 3KND_{max}^2 + K\mathbb{E}[\|\mathbf{A}\|^2] + 2K^2\|\mathbf{a}\|^2$$
$$+ 2(K\mathbf{X}_t\mathbf{a} - \sum_{k=0}^{K-1} \mathbb{E}[\mathbf{X}_{t+k}\mathbf{D}_{t+k}|\mathbf{Y}_t]$$
$$\leq 3KND_{max}^2 + K\mathbb{E}[\|\mathbf{A}\|^2] + 2K^2\|\mathbf{a}\|^2$$
$$2(K\mathbf{X}_t\mathbf{a} - \sum_{k=0}^{K-1} \mathbb{E}[\mathbf{X}_{t+k}\mathbf{D}_{t+k}^*|\mathbf{Y}_t])$$
$$+ 2\sum_{k=0}^{K-1} \mathbb{E}[\mathbf{X}_{t+k}(\mathbf{D}_{t+k}^* - \mathbf{D}_{t+k})|\mathbf{Y}_t]. \quad (52)$$

Using Lemmas 4, 6, 7, we can find an upper-bound for the

---

[10] Note that the sequence of $\epsilon_i$'s, $1 \leq i \leq 5$, here is different from the ones in Section V-A.

above, leading to the following for $\|\mathbf{X}_t\| \geq B$:[11]

$$\Delta(K)_t \leq 3KND_{max}^2 + KE[\|\mathbf{A}\|^2] + 2K^2\|\mathbf{a}\|^2$$

$$2K^2ND_{max}^2 + 2KD_{max}\sqrt{N}\epsilon_1\|\mathbf{X}_t\| \qquad (53)$$

$$- 2K\sum_{\mathbf{s}\in\mathcal{S}}\pi(\mathbf{s})(1 - \sum_{\mathbf{I}\in\mathcal{I}}\beta_{\mathbf{s},\mathbf{I}})\mathbf{X}_t\mathbf{D}^*(\mathbf{X}_t,\mathbf{s})$$

$$+ 2\min\left(KC_2 + K^2C_3 + K\sqrt{N}\epsilon_6\|\mathbf{X}_t\| + K\rho\alpha\delta^{-1}\|\mathbf{X}_t\|\right.$$

$$+ K\Upsilon(\mathbf{X}_t)\left(\zeta' + \frac{(1-\delta)}{\delta}\Psi_{\mathbf{Y}_t}^K\right),$$

$$\left.K^2C_3' + K\sqrt{N}\epsilon_6'\|\mathbf{X}_t\| + K\Upsilon(\mathbf{X}_t)(1-\Phi_{\mathbf{Y}_t}^K)\right), \quad (54)$$

where $C_2$, $C_3$, and $\epsilon_6$ are given in Lemma 6, and $C_3'$ and $\epsilon_6'$ are given in Lemma 7.

After a few simple algebraic steps, we obtain

$$\Delta(K)_t \leq KC_4 + K^2C_5$$

$$- 2K\left(1 - \max_{\mathbf{s}\in\mathcal{S}}\sum_{\mathbf{I}\in\mathcal{I}}\beta_{\mathbf{s},\mathbf{I}} -\right.$$

$$\min\left(\frac{\rho\alpha\|\mathbf{X}_t\|}{\delta\,\Upsilon(\mathbf{X}_t)} + \zeta' + \frac{1-\delta}{\delta}\Psi_{\mathbf{Y}_t}^K, 1 - \Phi_{\mathbf{Y}_t}^K\right)\Upsilon(\mathbf{X}_t)$$

$$+ 2K\sqrt{N}(\epsilon_1 D_{max} + \epsilon_6)\|\mathbf{X}_t\|, \quad (55)$$

where $C_4$ and $C_5$ are constants defined by $C_4 = 2C_2 + N(\tilde{A}_{max}^2 + 3D_{max}^2)$ and $C_5 = 2C_3 + 2\|\mathbf{a}\|^2 + 2ND_{max}^2$.

With our choices for $\epsilon_i$'s, $1 \leq i \leq 5$, and $K$, it is easy to verify that

$$\epsilon_1 D_{max} + \epsilon_6 < \frac{\nu}{N}\frac{\epsilon}{4}. \quad (56)$$

In addition, using the inequality in (20), the assumption that $K \geq K_{\epsilon_7}^{(\theta)}$ with $\epsilon_7 = \frac{\epsilon}{4}$, and the fact that $min(f(x)) = -max(-f(x))$, we can show that

$$1 - \min\left(\frac{\rho\alpha\|\mathbf{X}_t\|}{\delta\,\Upsilon(\mathbf{X}_t)} + \zeta' + \frac{1-\delta}{\delta}\Psi_{\mathbf{Y}_t}^K, 1 - \Phi_{\mathbf{Y}_t}^K\right)$$

$$\geq \max\left(1 - \frac{\sqrt{N}\rho\alpha}{\delta\nu} - \zeta' - \frac{1-\delta}{\delta}\Psi_{\mathbf{Y}_t}^K, \Phi_{\mathbf{Y}_t}^K\right)$$

$$> \theta - \epsilon_7 = \theta - \frac{\epsilon}{4}.$$

Using (20), (49), (55), and the previous inequality, we have

$$\Delta(K)_t \leq KC_4 + K^2C_5 - \frac{2\nu}{\sqrt{N}}K(\epsilon - \frac{\epsilon}{4} - \frac{N}{\nu}\epsilon_8)\|\mathbf{X}_t\|,$$

where $\epsilon_8 = (\epsilon_1 D_{max} + \epsilon_6)$. Using (56), for $\|\mathbf{X}_t\| \geq B$, we obtain

$$\Delta(K)_t \leq C_K - K\xi\|\mathbf{X}_t\|,$$

where $\xi = \frac{\nu\epsilon}{\sqrt{N}}$ and

$$C_K = KC_4 + K^2C_5. \quad (57)$$

Using the assumption that the second moments of the arrival process are finite as specified in (3), we can generalize the above inequality for all $\|\mathbf{X}_t\|$ as

$$\Delta(K)_t \leq C_K - K\xi\,\|\mathbf{X}_t\|\mathbf{1}_{\|\mathbf{X}_t\|\geq B} + C_{K,B}\mathbf{1}_{\|\mathbf{X}_t\|<B}$$

$$\leq -K\xi\,\|\mathbf{X}_t\|\mathbf{1}_{\|\mathbf{X}_t\|\geq B} + C_K + C_{K,B},$$

where $\mathbf{1}_{(\cdot)}$ is the indicator function and

$$C_{K,B} = 2KB\|\mathbf{a}\| + KE[\|\mathbf{A}\|^2] + K(K-1)\|\mathbf{a}\|^2$$

$$+ K^2ND_{max}^2. \quad (58)$$

Now, we take the expectation of $\Delta(K)_t$ with respect to the distribution of $\mathbf{Y}_t$, which leads to

$$\Delta'(K)_t \triangleq E[\|\mathbf{X}_{t+K}\|^2] - E[\|\mathbf{X}_t\|^2]$$

$$\leq -K\xi\,E[\,\|\mathbf{X}_t\|\mathbf{1}_{\|\mathbf{X}_t\|\geq B}\,] + C_K + C_{K,B}.$$

Considering the above inequality for times $i + jK$ for $i \in \{0,\ldots,K-1\}$ and $j \in \{0,\ldots,J-1\}$, and summing over $i$ and $j$, we obtain

$$\sum_{i=0}^{K-1}\sum_{j=0}^{J-1}\Delta'(K)_{i+jK} = \sum_{i=0}^{K-1}\left(E[\|\mathbf{X}_{i+JK}\|^2] - E[\|\mathbf{X}_i\|^2]\right)$$

$$\leq -K\xi\sum_{i=0}^{K-1}\sum_{j=0}^{J-1}E[\|\mathbf{X}_{i+jK}\|\mathbf{1}_{\|\mathbf{X}_{i+jK}\|\geq B}]$$

$$+ JK(C_K + C_{K,B}).$$

Since norm is a non-negative function, from the above, we have that

$$\sum_{i=0}^{K-1}\sum_{j=0}^{J-1}E[\|\mathbf{X}_{i+jK}\|\mathbf{1}_{\|\mathbf{X}_{i+jK}\|>B}]$$

$$\leq \frac{1}{K\xi}\sum_{i=0}^{K-1}E[\|\mathbf{X}_i\|^2] + \frac{J}{\xi}(C_K + C_{K,B}).$$

Using the fact that $E[\|\mathbf{X}_t\|\mathbf{1}_{\|\mathbf{X}_t\|<B}] < B$, and letting $t = i + jK$ and $T = JK - 1$, we obtain

$$\sum_{t=0}^{T}E[\|\mathbf{X}_t\|] \leq (T+1)B + \frac{1}{K\xi}\sum_{i=0}^{K-1}E[\|\mathbf{X}_i\|^2]$$

$$+ \frac{J}{\xi}(C_K + C_{K,B}). \quad (59)$$

Since the first and second moments of the arrival process are finite, for a fixed $K$, the summation on the righthand side of (59) can be bounded by an appropriate constant $C_6$. Hence, we must have $\sum_{i=0}^{K-1}E[\|\mathbf{X}_i\|^2] < C_6$. Therefore, we have

$$\sum_{t=0}^{T}E[\|\mathbf{X}_t\|] \leq (T+1)B + \frac{1}{K\xi}C_6 + \frac{J}{\xi}(C_K + C_{K,B}).$$

Finally, dividing by $T + 1$ and letting $T \to \infty$, by assuming $J \to \infty$, we obtain

$$\limsup_{T\to\infty}\frac{1}{T+1}\sum_{t=0}^{T}E[\|\mathbf{X}_t\|] \leq B + \frac{C_K + C_{K,B}}{\xi K} < \infty, \quad (60)$$

which completes the proof of part (b) of the theorem. $\qquad\square$

*Proof of Theorem 2.* To start, first note that $\sum_{l=1}^{N}X_l(t) \leq \sqrt{N}\|\mathbf{X}_t\|$. We can use this inequality, the one in (60), and equality $\xi = \frac{\nu\epsilon}{\sqrt{N}}$ along with the definitions for $C_K$ and $C_{K,B}$, given in (57) and (58), respectively, to show that

$$\limsup_{T\to\infty}\frac{1}{T+1}\sum_{t=0}^{T}\left[\sum_{l=1}^{N}E[X_l(t)]\right] < B(\sqrt{N} + 2\frac{N}{\nu\epsilon}\|\mathbf{a}\|)$$

$$+ \frac{N}{\nu\epsilon}\left(2N\tilde{A}_{max}^2 + 11\delta^{-1}ND_{max}^2 + 16\,\delta^{-1}ND_{max}^2K\right.$$

$$\left.+ 4\delta^{-1}\rho\alpha\sqrt{N}(\tilde{A}_{max} + D_{max})K\right), \quad (61)$$

where according to the proof in part (a), $K = K_\epsilon$ and $B = B_\epsilon^K$; see (50) and (51), respectively. To obtain the above inequality, we have also used the fact that if $\mathbf{a} \in \Gamma$, then $a_l < D_{max}$, $1 \leq l \leq N$, and hence, $\|\mathbf{a}\| < \sqrt{N} D_{max}$.

---

[11]Note that both Lemma 6 and Lemma 7 provide an upper-bound for the last term in (52), and we use the minimum upper-bound in our analysis.

Note that $K$ depends on $N$ and $\delta$. Therefore, to complete the proof, it suffices to show that the expression in the large brackets in (61) is $\mathcal{O}(KN\delta^{-1})$ with respect to variables $\delta$ and $N$. This easily follows since by (23), $\sqrt{N} \rho \alpha \delta^{-1} \nu^{-1} \leq 1$, completing the proof. $\qquad \square$

## APPENDIX
## LEMMAS

Here, we often use subscripts to denote time dependencies. We use $\mathbf{1}$ to represent a vector with all its elements equal to one and use $\mathbf{1}_e$ to denote the indicator function for the event $e$. Where required, we use underlines to show realizations of r.v.'s, e.g., in our analysis, $\underline{\mathbf{X}}_{t+m}$ denotes one realization of the random vector $\mathbf{X}_{t+m}$.

**Lemma 2.** *For all values of $t$, $\mathbf{s} \in \mathcal{S}$, and $m \geq 0$, we have that*
$$\mathbb{E}\big[\mathbf{X}_{t+m+1}\big(\mathbf{D}^*(\mathbf{X}_{t+m+1}, \mathbf{s}) - \mathbf{D}^*(\mathbf{X}_{t+m}, \mathbf{s})\big)|\mathbf{Y}_t\big] < C_1,$$
*where $C_1 = \sqrt{N} D_{max} \|\mathbf{a}\| + N D_{max}^2$.*

*Proof.* By the definition of $\mathbf{D}^*(\mathbf{X}, \mathbf{s})$, as provided in (10), we have
$$\mathbf{X}_{t+m+1}\big(\mathbf{D}^*(\mathbf{X}_{t+m+1}, \mathbf{s}) - \mathbf{D}^*(\mathbf{X}_{t+m}, \mathbf{s})\big)$$
$$= \mathbf{X}_{t+m}\big(\mathbf{D}^*(\mathbf{X}_{t+m+1}, \mathbf{s}) - \mathbf{D}^*(\mathbf{X}_{t+m}, \mathbf{s})\big) +$$
$$\big(\mathbf{X}_{t+m+1} - \mathbf{X}_{t+m}\big)\big(\mathbf{D}^*(\mathbf{X}_{t+m+1}, \mathbf{s}) - \mathbf{D}^*(\mathbf{X}_{t+m}, \mathbf{s})\big)$$
$$\leq \big(\mathbf{X}_{t+m+1} - \mathbf{X}_{t+m}\big)\big(\mathbf{D}^*(\mathbf{X}_{t+m+1}, \mathbf{s}) - \mathbf{D}^*(\mathbf{X}_{t+m}, \mathbf{s})\big)$$
$$= \big(\mathbf{A}_{t+m} - (\mathbf{D}_{t+m} - \mathbf{U}_{t+m})\big)$$
$$\big(\mathbf{D}^*(\mathbf{X}_{t+m+1}, \mathbf{s}) - \mathbf{D}^*(\mathbf{X}_{t+m}, \mathbf{s})\big)$$
$$\leq \mathbf{A}_{t+m}\mathbf{D}^*(\mathbf{X}_{t+m+1}, \mathbf{s}) + \mathbf{D}_{t+m}\mathbf{D}^*(\mathbf{X}_{t+m}, \mathbf{s})$$
$$\leq D_{max}\mathbf{A}_{t+m}.\mathbf{1} + N D_{max}^2,$$
where the last inequality follows from the assumption that transmission rates are bounded by $D_{max}$. Taking the expected value of both sides and noting that $\mathbf{A}_{t+m}$ is independent of $\mathbf{Y}_t$, proves the Lemma with the given $C_1$. $\qquad \square$

**Lemma 3.** *For any two queue-length vectors $\mathbf{X}_1$ and $\mathbf{X}_2$, and $(\mathbf{s}_1, \mathbf{s}_2) \in \mathcal{S} \times \mathcal{S}$, the following holds:*
$$\mathbf{X}_2\big(\mathbf{D}^*(\mathbf{X}_2, \mathbf{s}_2) - \mathbf{D}^*(\mathbf{X}_2, \mathbf{s}_1)\big)$$
$$-\mathbf{X}_1\big(\mathbf{D}^*(\mathbf{X}_1, \mathbf{s}_2) - \mathbf{D}^*(\mathbf{X}_1, \mathbf{s}_1)\big)$$
$$\leq (\mathbf{X}_2 - \mathbf{X}_1)\big(\mathbf{D}^*(\mathbf{X}_2, \mathbf{s}_2) - \mathbf{D}^*(\mathbf{X}_1, \mathbf{s}_1)\big).$$

*Proof.* The proof easily follows from the definition of $\mathbf{D}^*(\mathbf{X}, \mathbf{s})$ given in (10). $\qquad \square$

**Lemma 4.** *Suppose $\mathbf{a}$ is given by (27), where non-negative constants $\beta_{\mathbf{s}, \mathbf{I}}$'s satisfy $\sum_{\mathbf{I} \in \mathcal{I}} \beta_{\mathbf{s}, \mathbf{I}} \leq 1$ for all $\mathbf{s} \in \mathcal{S}$. For any positive $\epsilon_1$, if $K \geq K_{1,\epsilon_1}$, then we have*
$$\Delta_{\mathbf{X}_t} \triangleq K\mathbf{X}_t\mathbf{a} - \sum_{k=0}^{K-1} \mathbb{E}[\mathbf{X}_{t+k}\mathbf{D}^*_{t+k}|\mathbf{Y}_t]$$
$$\leq K^2 N D_{max}^2 + K D_{max}\sqrt{N}\epsilon_1\|\mathbf{X}_t\|$$
$$- K \sum_{\mathbf{s}} \pi(\mathbf{s})(1 - \sum_{\mathbf{I} \in \mathcal{I}} \beta_{\mathbf{s}, \mathbf{I}})\mathbf{X}_t\mathbf{D}^*(\mathbf{X}_t, \mathbf{s}),$$

*where $K_{1,\epsilon_1}$ is defined in Section III-B.*

*Proof.* First, note that the following holds using the definition for $\mathbf{D}^*(\mathbf{X}, \mathbf{s})$ and the assumption that $D_{max}$ is the global upper bound for individual rates.
$$\mathbf{X}_{t+k}\mathbf{D}^*_{t+k} = \max_{\mathbf{I} \in \mathcal{I}} \mathbf{X}_{t+k}\mathbf{D}(\mathbf{s}_{t+k}, \mathbf{I})$$
$$\geq \max_{\mathbf{I} \in \mathcal{I}} \mathbf{X}_t\mathbf{D}(\mathbf{s}_{t+k}, \mathbf{I}) - \max_{\mathbf{I} \in \mathcal{I}} \big(\sum_{i=0}^{k-1} \mathbf{D}_{t+i}\mathbf{D}(\mathbf{s}_{t+k}, \mathbf{I})\big)$$
$$\geq \max_{\mathbf{I} \in \mathcal{I}} \mathbf{X}_t\mathbf{D}(\mathbf{s}_{t+k}, \mathbf{I}) - k N D_{max}^2. \qquad (62)$$

In addition, for a given $\mathbf{X}$, by the assumption in the lemma and the definition for $\Upsilon(\mathbf{X})$ given in (16), we have
$$\mathbf{Xa} - \Upsilon(\mathbf{X})$$
$$= \mathbf{X}\sum_{\mathbf{s} \in \mathcal{S}} \pi(\mathbf{s})\sum_{\mathbf{I} \in \mathcal{I}} \beta_{\mathbf{s}, \mathbf{I}}\mathbf{D}(\mathbf{s}, \mathbf{I}) - \sum_{\mathbf{s} \in \mathcal{S}} \pi(\mathbf{s})\mathbf{X}\mathbf{D}^*(\mathbf{X}, \mathbf{s})$$
$$= \sum_{\mathbf{s} \in \mathcal{S}} \pi(\mathbf{s})\sum_{\mathbf{I} \in \mathcal{I}} \beta_{\mathbf{s}, \mathbf{I}}\big(\mathbf{XD}(\mathbf{s}, \mathbf{I}) - \mathbf{XD}^*(\mathbf{X}, \mathbf{s})\big)$$
$$- \sum_{\mathbf{s} \in \mathcal{S}} \pi(\mathbf{s})\big(1 - \sum_{\mathbf{I} \in \mathcal{I}} \beta_{\mathbf{s}, \mathbf{I}}\big)\mathbf{XD}^*(\mathbf{X}, \mathbf{s})$$
$$\leq -\sum_{\mathbf{s} \in \mathcal{S}} \pi(\mathbf{s})\big(1 - \sum_{\mathbf{I} \in \mathcal{I}} \beta_{\mathbf{s}, \mathbf{I}}\big)\mathbf{XD}^*(\mathbf{X}, \mathbf{s}), \qquad (63)$$

where the last inequality follows from the definition of $\mathbf{D}^*(\mathbf{X}, \mathbf{s})$ implying that $\mathbf{XD}(\mathbf{s}, \mathbf{I}) \leq \mathbf{XD}^*(\mathbf{X}, \mathbf{s})$ for all $\mathbf{I} \in \mathcal{I}$.

Now, we can use (62) to show that
$$\Delta_{\mathbf{X}_t} \leq K^2 N D_{max}^2 + K\mathbf{X}_t\mathbf{a}$$
$$- \sum_{k=0}^{K-1} \mathbb{E}\Big[\mathbf{X}_t \sum_{\mathbf{s} \in \mathcal{S}} \mathbf{1}_{\mathbf{s}(t+k)=\mathbf{s}}\mathbf{D}^*(\mathbf{X}_t, \mathbf{s})\big|\mathbf{Y}_t\Big]$$
$$= K^2 N D_{max}^2 + K\mathbf{X}_t\mathbf{a}$$
$$- \sum_{\mathbf{s} \in \mathcal{S}} \mathbf{X}_t\mathbb{E}\big[\sum_{k=0}^{K-1} \mathbf{1}_{\mathbf{s}(t+k)=\mathbf{s}}|\mathbf{Y}_t\big]\mathbf{D}^*(\mathbf{X}_t, \mathbf{s})$$
$$+ K\sum_{\mathbf{s} \in \mathcal{S}} \pi(\mathbf{s})\mathbf{X}_t\mathbf{D}^*(\mathbf{X}_t, \mathbf{s}) - K\sum_{\mathbf{s} \in \mathcal{S}} \pi(\mathbf{s})\mathbf{X}_t\mathbf{D}^*(\mathbf{X}_t, \mathbf{s})$$
$$= K^2 N D_{max}^2$$
$$+ K\sum_{\mathbf{s} \in \mathcal{S}} \big(\pi(\mathbf{s}) - \mathbb{E}\big[\frac{1}{K}\sum_{k=0}^{K-1} \mathbf{1}_{\mathbf{s}(t+k)=\mathbf{s}}|\mathbf{Y}_t\big]\big)\mathbf{X}_t\mathbf{D}^*(\mathbf{X}_t, \mathbf{s})$$
$$+ K\big(\mathbf{X}_t\mathbf{a} - \Upsilon(\mathbf{X}_t)\big),$$

where in the last equality, we have used (16).

When $K \geq K_{1,\epsilon_1}$, the inequality (6) holds for $\epsilon = \epsilon_1$. Using this, inequality (63), and that transmission rates are bounded by $D_{max}$, we obtain the inequality in the lemma for $K \geq K_{1,\epsilon_1}$, as required. $\qquad \square$

**Lemma 5.** *Suppose arbitrary positive integer $K$ and arbitrary positive real numbers $\epsilon_2$ and $\epsilon_3$ are given. Let $\tilde{A}_{\epsilon_2} = \sqrt{N} D_{max} + A_{\epsilon_2}$, where $A_{\epsilon_2}$ is defined by Fact 1, and let $B_{\epsilon_2, \epsilon_3}^K = \max(B_{1,\epsilon_3}^C, B_{2,\epsilon_3}^C)$, where $C = K\tilde{A}_{\epsilon_2}$ and $B_{1,\epsilon_3}^C$ and $B_{2,\epsilon_3}^C$ are defined by Property 2 and Fact 2, respectively. Then,* *the following inequality holds if $\|\mathbf{X}_t\| \geq B_{\epsilon_2,\epsilon_3}^K$:*

$$\left| \sum_{m=0}^{K-2} \mathbb{E}\Big[\mathbf{X}_t\big(\mathbf{D}_{t+m} - (1-\rho)\mathbf{D}'_{t+m}\big)\big|\mathbf{Y}_t\Big] \right.$$
$$\left. - \sum_{m=0}^{K-2} \bar{\mathbb{E}}_{\mathbf{X}_t}\Big[\mathbf{X}_t\big(\mathbf{D}_{t+m} - (1-\rho)\mathbf{D}'_{t+m}\big)\big|\mathbf{Y}_t\Big] \right|$$
$$\leq K D_{max}\sqrt{N}\Big(2K\epsilon_2 + \big((1+2\epsilon_3)^{2K} - 1\big)\Big)\|\mathbf{X}_t\|.$$

*Proof.* Recall that the notation $\bar{\mathbb{E}}_{\mathbf{X}_t}$ represents the expectation assuming that the policy does not have updated queue information after time $t$. Hence, while in the first summation in the lemma, we are taking a typical expectation considering all dynamics of the system, in the second summation, we are taking the expectation assuming that at any time $t+m$, $m \geq 0$, the scheduling policy updates the schedule vector $\mathbf{I}_{t+m}$ using the old queue information at time $t$, and therefore, by setting $\mathbf{X}_{t+m} = \mathbf{X}_t$, $m \geq 0$. In the following, to distinguish a particular realization of a random variable, we use underlines, and in the subscripts, we remove dependencies on time. For instance, $\underline{\mathbf{I}}_t^r$ represents one realization of $\mathbf{I}_{t+i}^r$.

For $K \geq 1$, let $\mathbf{X}_t^K = \{\mathbf{X}_{t+i}\}_{i=1}^K$, $\mathbf{s}_t^K = \{\mathbf{s}_{t+i}\}_{i=1}^K$, $\mathbf{I}_t^{r,K} = \{\mathbf{I}_{t+i}^r\}_{i=1}^K$, and $\mathbf{I}_t^K = \{\mathbf{I}_{t+i}\}_{i=1}^K$. For $k \geq 1$, the following equalities follow easily from the update rule properties and using the fact that the arrival process is an i.i.d process and independent of the channel process.

$$p(\mathbf{X}_{t+k}|\mathbf{X}_t^{k-1}, \mathbf{s}_t^{k-1}, \mathbf{I}_t^{r,k-1}, \mathbf{I}_t^{k-1}, \mathbf{Y}_t)$$
$$= p(\mathbf{X}_{t+k}|\mathbf{X}_{t+k-1}, \mathbf{s}_{t+k-1}, \mathbf{I}_{t+k-1}),$$
$$p(\mathbf{s}_{t+k}|\mathbf{X}_t^k, \mathbf{s}_t^{k-1}, \mathbf{I}_t^{r,k-1}, \mathbf{I}_t^{k-1}, \mathbf{Y}_t)$$
$$= p(\mathbf{s}_{t+k}|\mathbf{s}_t^{k-1}, \mathbf{Y}_t),$$
$$p(\mathbf{I}_{t+k}^r|\mathbf{X}_t^k, \mathbf{s}_t^k, \mathbf{I}_t^{r,k-1}, \mathbf{I}_t^{k-1}, \mathbf{Y}_t)$$
$$= p(\mathbf{I}_{t+k}^r|\mathbf{X}_{t+k}, \mathbf{s}_{t+k}),$$
$$p(\mathbf{I}_{t+k}|\mathbf{X}_t^k, \mathbf{s}_t^k, \mathbf{I}_t^{r,K}, \mathbf{I}_t^{k-1}, \mathbf{Y}_t)$$
$$= p(\mathbf{I}_{t+k}|\mathbf{X}_{t+k}, \mathbf{s}_{t+k}, \mathbf{I}_{t+k}^r, \mathbf{I}_{t+k-1}).$$

Let $\mathbf{W}_{t+i}$, $i \geq 1$, be defined as

$$\mathbf{W}_{t+i} = (\mathbf{X}_{t+i}, \mathbf{s}_{t+i}, \mathbf{I}_{t+i}^r, \mathbf{I}_{t+i}),$$

and $\mathbf{W}_t^K$ be $\{\mathbf{W}_{t+i}\}_{i=1}^K$. Hence, $\mathbf{W}_{t+i}$ contains all r.v.'s associated with time $t+i$, and $\mathbf{W}_t^K$ is the collection of $\mathbf{W}_{t+i}$'s over $K$ time-slots. We also use the same variable $\mathbf{W}_t^K$ to represent $(\mathbf{X}_t^K, \mathbf{s}_t^K, \mathbf{I}_t^{r,K}, \mathbf{I}_t^K)$ that essentially contains the same set of r.v.'s as does $\{\mathbf{W}_{t+i}\}_{i=1}^K$. Using conditional probabilities and the above equalities, we have

$$p(\mathbf{W}_t^K|\mathbf{Y}_t) =$$
$$\prod_{i=1}^K \Big( p(\mathbf{X}_{t+i}|\mathbf{X}_{t+i-1}, \mathbf{s}_{t+k-1}, \mathbf{I}_{t+i-1}) p(\mathbf{s}_{t+i}|\mathbf{s}_t^{i-1}, \mathbf{Y}_t)$$
$$p(\mathbf{I}_{t+i}^r|\mathbf{X}_{t+i}, \mathbf{s}_{t+i}) p(\mathbf{I}_{t+i}|\mathbf{X}_{t+i}, \mathbf{s}_{t+i}, \mathbf{I}_{t+i}^r, \mathbf{I}_{t+i-1})\Big). \quad (64)$$

Similarly, for the case where the policy does not have updated queue information, let $\hat{\mathbf{I}}_t^{r,K} = \{\hat{\mathbf{I}}_{t+i}^r\}_{i=1}^K$, and $\hat{\mathbf{I}}_t^K = \{\hat{\mathbf{I}}_{t+i}\}_{i=1}^K$, where $\hat{\mathbf{I}}_{t+i}^r$ and $\hat{\mathbf{I}}_{t+i}$ are the candidate schedule and the updated schedule at time $t+i$, respectively, based on the assumption that after time $t$, old queue information at time $t$ is used by the policy. Let $\hat{\mathbf{W}}_{t+i}$ be

$$\hat{\mathbf{W}}_{t+i} = (\mathbf{s}_{t+i}, \hat{\mathbf{I}}_{t+i}^r, \hat{\mathbf{I}}_{t+i}),$$

and $\hat{\mathbf{W}}_t^K = \{\hat{\mathbf{W}}_{t+i}\}_{i=1}^K$. We use the same variable $\hat{\mathbf{W}}_t^K$ to represent $(\mathbf{s}_t^K, \hat{\mathbf{I}}_t^{r,K}, \hat{\mathbf{I}}_t^K)$. Using the same approach to simplify $p(\mathbf{W}_t^K|\mathbf{Y}_t)$, we can show that

$$\bar{p}_{\mathbf{X}_t}(\hat{\mathbf{W}}_t^K|\mathbf{Y}_t) = \prod_{i=1}^K \Big( p(\mathbf{s}_{t+i}|\mathbf{s}_t^{i-1}, \mathbf{Y}_t) p(\hat{\mathbf{I}}_{t+i}^r|\mathbf{X}_t, \mathbf{s}_{t+i})$$
$$p(\hat{\mathbf{I}}_{t+i}|\mathbf{X}_t, \mathbf{s}_{t+i}, \hat{\mathbf{I}}_{t+i}^r, \hat{\mathbf{I}}_{t+i-1})\Big), \quad (65)$$

where the notation $\bar{p}_{\mathbf{X}_t}(\cdot)$, as defined in Section V-A, is used to emphasize that $\hat{\mathbf{W}}_t^K$ is a set of r.v.'s with the assumption that the policy does not have the updated queue information after time $t$.

To continue, let $\mathbf{X}_t^K \triangleq (\mathbf{X}_{t+1}, \ldots, \mathbf{X}_{t+K})$ and

$$\mathcal{R}(\mathbf{X}_t, K, \tilde{A}_{\epsilon_2}) = \{\underline{\mathbf{X}}^K; \|\underline{\mathbf{X}}_1 - \mathbf{X}_t\| \leq \tilde{A}_{\epsilon_2},$$
$$\|\underline{\mathbf{X}}_i - \underline{\mathbf{X}}_{i-1}\| \leq \tilde{A}_{\epsilon_2}, i = 2, \ldots, K\}.$$

Note that, as mentioned in the beginning of the proof, $\underline{\mathbf{X}}_i$ represents one realization of $\mathbf{X}_{t+i}$. By assumption, $\tilde{A}_{\epsilon_2} - \sqrt{N}D_{max} = A_{\epsilon_2}$, and hence, by Fact 1, regardless of $\mathbf{X}_{t+i-1}$, $\mathbf{s}_{t+i-1}$, and $\mathbf{I}_{t+i-1}$, we have

$$p(\|\mathbf{X}_{t+i} - \mathbf{X}_{t+i-1}\| > \tilde{A}_{\epsilon_2}\,|\, \mathbf{X}_{t+i-1}, \mathbf{s}_{t+i-1}, \mathbf{I}_{t+i-1})$$
$$\leq p(\|\mathbf{A}_{t+i-1}\| > \tilde{A}_{\epsilon_2} - \sqrt{N}D_{max})$$
$$< \epsilon_2. \quad (66)$$

Using the above inequality and the union bound, we have

$$p(\mathbf{X}_t^K \notin \mathcal{R}(\mathbf{X}_t, K, \tilde{A}_{\epsilon_2})|\mathbf{Y}_t) < K\epsilon_2. \quad (67)$$

By definition, $\mathbf{X}_t^K \in \mathcal{R}(\mathbf{X}_t, K, \tilde{A}_{\epsilon_2})$ implies $\|\mathbf{X}_{t+i} - \mathbf{X}_t\| \leq K\tilde{A}_{\epsilon_2}$, for $1 \leq i \leq K$. Therefore, if $\mathbf{X}_t^K \in \mathcal{R}(\mathbf{X}_t, K, \tilde{A}_{\epsilon_2})$ and $\|\mathbf{X}_t\| \geq B_{\epsilon_2,\epsilon_3}^C = \max(B_{1,\epsilon_3}^C, B_{2,\epsilon_3}^C)$, where $C = K\tilde{A}_{\epsilon_2}$, by Property 2, we have

$$\sum_{\mathbf{I} \in \mathcal{I}} \left| p(\mathbf{I}_{t+i}^r = \mathbf{I}|\mathbf{X}_{t+i}, \mathbf{s}_{t+i}) - p(\hat{\mathbf{I}}_{t+i}^r = \mathbf{I}|\mathbf{X}_t, \mathbf{s}_{t+i}) \right| < \epsilon_3. \quad (68)$$

for all $\mathbf{I} \in \mathcal{I}$ and $\mathbf{s}_{t+i} \in \mathcal{S}$. Similarly, assuming–for a particular realization–$\mathbf{I}_{t+i}^r = \hat{\mathbf{I}}_{t+i}^r = \underline{\mathbf{I}}_i^r$ and $\mathbf{I}_{t+i-1} = \hat{\mathbf{I}}_{t+i-1} = \underline{\mathbf{I}}_{i-1}$, by Fact 2, we have

$$\left| p(\mathbf{I}_{t+i} = \mathbf{I}|\mathbf{X}_{t+i}, \mathbf{s}_{t+i}, \mathbf{I}_{t+i}^r, \mathbf{I}_{t+i-1}) \right.$$
$$\left. - p(\hat{\mathbf{I}}_{t+i} = \mathbf{I}|\mathbf{X}_t, \mathbf{s}_{t+i}, \hat{\mathbf{I}}_{t+i}^r, \hat{\mathbf{I}}_{t+i-1}) \right| < \epsilon_3, \quad (69)$$

for $\mathbf{I} \in \{\underline{\mathbf{I}}_i^r, \underline{\mathbf{I}}_{i-1}\}$, all $\mathbf{s} \in \mathcal{S}$, and all $\underline{\mathbf{I}}_i^r$ and $\underline{\mathbf{I}}_{i-1}$ in $\mathcal{I}$.

We are now ready to consider the summation terms in the lemma. First, for a given $\underline{\mathbf{s}}^K = \{\underline{\mathbf{s}}_i\}_{i=1}^K$ and $\underline{\mathbf{I}}^K = \{\underline{\mathbf{I}}_i\}_{i=1}^K$, define the function $g(\underline{\mathbf{s}}^K, \underline{\mathbf{I}}^K)$ as

$$g(\underline{\mathbf{s}}^K, \underline{\mathbf{I}}^K) = \sum_{i=0}^{K-1} \mathbf{X}_t\big(\mathbf{D}(\mathbf{s}_i, \mathbf{I}_i) - (1-\rho)\mathbf{D}(\mathbf{s}_{i+1}, \mathbf{I}_i)\big). \quad (70)$$

Using this function, we define the r.v. $g(\mathbf{W}_t^K)$ as

$$g(\mathbf{W}_t^K) = g(\mathbf{I}_t^K, \mathbf{s}_t^K) = \sum_{m=0}^{K-1} \mathbf{X}_t\Big(\mathbf{D}_{t+m} - (1-\rho)\mathbf{D}'_{t+m}\Big)$$
$$= \sum_{m=0}^{K-1} \mathbf{X}_t\Big(\mathbf{D}(\mathbf{s}_{t+m}, \mathbf{I}_{t+m}) - (1-\rho)\mathbf{D}(\mathbf{s}_{t+m+1}, \mathbf{I}_{t+m})\Big). \quad (71)$$

Setting $\underline{\mathbf{W}}^K$ as one realization of $\mathbf{W}_t^K$, which contains the realization sequences $\underline{\mathbf{X}}^K$, $\underline{\mathbf{s}}^K$, $\underline{\mathbf{I}}^{r,K}$, and $\underline{\mathbf{I}}^K$, where, e.g., $\underline{\mathbf{I}}^{r,K}$ represents one realization of $\mathbf{I}_t^{r,K}$, we have

$$\mathbb{E}[g(\mathbf{W}_t^K)|\mathbf{Y}_t]$$
$$= \sum_{\underline{\mathbf{W}}^K} p(\mathbf{W}_t^K = \underline{\mathbf{W}}^K|\mathbf{Y}_t)g(\underline{\mathbf{W}}^K)$$
$$= \sum_{\underline{\mathbf{W}}^K : \underline{\mathbf{X}}^K \notin \mathcal{R}(\mathbf{X}_t, K, \tilde{A}_{\epsilon_2})} p(\mathbf{W}_t^K = \underline{\mathbf{W}}^K|\mathbf{Y}_t)g(\underline{\mathbf{W}}^K) \quad (72)$$
$$+ \sum_{\underline{\mathbf{W}}^K : \underline{\mathbf{X}}^K \in \mathcal{R}(\mathbf{X}_t, K, \tilde{A}_{\epsilon_2})} p(\mathbf{W}_t^K = \underline{\mathbf{W}}^K|\mathbf{Y}_t)g(\underline{\mathbf{W}}^K). \quad (73)$$

In above, we have separated two cases by considering whether or not $\mathbf{X}_t^K \in \mathcal{R}(\mathbf{X}_t, K, \tilde{A}_{\epsilon_2})$. We have also used under-lined variables to highlight realizations of the their associated random variables. Since for all $\underline{\mathbf{W}}^K$

$$|g(\underline{\mathbf{W}}^K)| < KD_{max}\sqrt{N}\|\mathbf{X}_t\|, \quad (74)$$

by (67), we have

$$(72) \leq K^2 D_{max}\sqrt{N}\epsilon_2\|\mathbf{X}_t\|. \quad (75)$$

To simplify (73), note that when $\mathbf{X}_t^K \in \mathcal{R}(\mathbf{X}_t, K, \tilde{A}_{\epsilon_2})$ and $\|\mathbf{X}_t\| \geq B_{\epsilon_2,\epsilon_3}^K$, inequalities (68) and (69) hold. Using these inequalities and (64), we see that for each realization $\underline{\mathbf{W}}^K$ of the random vector $\mathbf{W}_t^K$, we have

$$p(\mathbf{W}_t^K = \underline{\mathbf{W}}^K|\mathbf{Y}_t) = \prod_{i=1}^{K} \Bigg[$$
$$p\Big(\mathbf{X}_{t+i} = \underline{\mathbf{X}}_i|\mathbf{X}_{t+i-1} = \underline{\mathbf{X}}_{i-1}, \mathbf{s}_{t+i-1} = \underline{\mathbf{s}}_{i-1},$$
$$\mathbf{I}_{t+i-1} = \underline{\mathbf{I}}_{i-1}\Big)$$
$$p\Big(\mathbf{s}_{t+i} = \underline{\mathbf{s}}_i|\mathbf{s}_t^{i-1} = \underline{\mathbf{s}}^{i-1}, \mathbf{Y}_t\Big)$$
$$\Big(p(\hat{\mathbf{I}}_{t+i}^r = \underline{\mathbf{I}}_i^r|\mathbf{X}_t, \mathbf{s}_{t+i} = \underline{\mathbf{s}}_i) + \xi_{2i-1}(\underline{\mathbf{I}}_i^r)\Big)$$
$$\Big(p(\hat{\mathbf{I}}_{t+i} = \underline{\mathbf{I}}_i|\mathbf{X}_t, \mathbf{s}_{t+i} = \underline{\mathbf{s}}_i, \hat{\mathbf{I}}_{t+i}^r = \underline{\mathbf{I}}_i^r, \hat{\mathbf{I}}_{t+i-1} = \underline{\mathbf{I}}_{i-1}$$
$$+ \xi_{2i}(\underline{\mathbf{I}}_i))\Bigg], \quad (76)$$

where $\xi_{2i-1}(\underline{\mathbf{I}}_i^r)$ and $\xi_{2i}(\underline{\mathbf{I}}_i)$, $1 \leq i \leq K$, that are specified as functions of $\underline{\mathbf{I}}_i^r$ and $\underline{\mathbf{I}}_i$, respectively, in general depend on $\underline{\mathbf{W}}^K$ with the property that $\sum_{\underline{\mathbf{I}}_i^r}|\xi_{2i-1}(\underline{\mathbf{I}}_i^r)| < \epsilon_3$, and $|\xi_{2i}(\underline{\mathbf{I}}_i)| < \epsilon_3$.

Considering (76), we see that there are $2^{2K}$ terms in the expression for $p(\mathbf{W}_t^K = \underline{\mathbf{W}}^K|\mathbf{Y}_t)$. Each term is obtained by including either the error terms $\xi_{2i-1}$ and $\xi_{2i}$ or their corresponding probability expressions, for all $i$, where $1 \leq i \leq K$. This implies that the summation in (73) can be decomposed into $2^{2K}$ sub-summations. The main sub-summation $\Sigma_m$ is the one where no error is present. Using (65) and (76), we have

$$\Sigma_m = \sum_{\underline{\mathbf{W}}^K = (\underline{\mathbf{s}}^K, \underline{\mathbf{I}}^r, K, \underline{\mathbf{I}}^K)} \bar{p}_{\mathbf{X}_t}(\hat{\mathbf{W}}_t^K = \underline{\hat{\mathbf{W}}}^K|\mathbf{Y}_t)g(\underline{\mathbf{I}}^K, \underline{\mathbf{s}}^K)$$
$$\sum_{\underline{\mathbf{X}}^K \in \mathcal{R}(\mathbf{X}_t, K, \tilde{A}_{\epsilon_2})} \prod_{i=1}^{K} \Big(p(\mathbf{X}_{t+i} = \underline{\mathbf{X}}_i|\mathbf{X}_{t+i-1} = \underline{\mathbf{X}}_{i-1},$$
$$\mathbf{s}_{t+i-1} = \underline{\mathbf{s}}_{i-1}, \mathbf{I}_{t+i-1} = \underline{\mathbf{I}}_{i-1})\Big). \quad (77)$$

Note that in above the summations are nested. By the definition for $\bar{\mathbb{E}}_{\mathbf{X}_t}[\cdot]$, we have

$$\sum_{\underline{\mathbf{W}}^K} \bar{p}_{\mathbf{X}_t}(\hat{\mathbf{W}}_t^K = \underline{\hat{\mathbf{W}}}^K|\mathbf{Y}_t)g(\underline{\mathbf{I}}^K, \underline{\mathbf{s}}^K) = \bar{\mathbb{E}}_{\mathbf{X}_t}[g(\mathbf{I}_t^K, \mathbf{s}_t^K)|\mathbf{Y}_t].$$

This equality along with the inequalities in (66) and (74) yields

$$|\Sigma_m - \bar{\mathbb{E}}_{\mathbf{X}_t}[g(\mathbf{I}^K, \mathbf{s}^K)|\mathbf{Y}_t]|$$
$$< K\sqrt{N}\|\mathbf{X}_t\|D_{max}(1 - (1 - \epsilon_2)^K)$$
$$< K^2\epsilon_2 D_{max}\sqrt{N}\|\mathbf{X}_t\|. \quad (78)$$

Now, we consider the effect of errors. Let $\Sigma_{k_1,k_2}^e$ be the sub-summation corresponding to the term with $k_1$ errors of the form $\xi_{2i_j-1}$, where $i_1 < i_2 < \ldots, < i_{k_1}$, and with $k_2$ errors of the form form $\xi_{2l_j}$, where $l_1 < l_2 < \ldots, < l_{k_2}$. Using (74) and that $|\xi_{2i}(\underline{\mathbf{I}}_i)| < \epsilon_3$, we have that

$$|\Sigma_e| \leq$$
$$\sum_{\underline{\mathbf{s}}_1 \in \mathcal{S}} p(\mathbf{s}_{t+1}|\mathbf{Y}_t) \sum_{\underline{\mathbf{I}}_1^r \in \mathcal{I}} p(\hat{\mathbf{I}}_{t+1}^r|\mathbf{X}_t, \mathbf{s}_{t+1})$$
$$\sum_{\underline{\mathbf{I}}_1 \in \{\underline{\mathbf{I}}_1^r, \mathbf{I}_t\}} p(\hat{\mathbf{I}}_{t+1}|\mathbf{X}_t, \mathbf{s}_{t+1}, \hat{\mathbf{I}}_{t+1}^r, \mathbf{I}_t) \cdots \sum_{\underline{\mathbf{s}}_i \in \mathcal{S}} p(\mathbf{s}_{t+i}|\mathbf{s}_t^{i-1}, \mathbf{Y}_t) \cdots$$
$$\sum_{\underline{\mathbf{I}}_{i_1} \in \mathcal{I}} \xi_{2i_1-1}(\underline{\mathbf{I}}_{i_1}^r) \cdots \sum_{\underline{\mathbf{I}}_l \in \{\underline{\mathbf{I}}_l^r, \underline{\mathbf{I}}_{l-1}\}, \ l < l_1} p(\hat{\mathbf{I}}_{t+l}|\mathbf{X}_t, \mathbf{s}_{t+l}, \hat{\mathbf{I}}_{t+l}^r, \hat{\mathbf{I}}_{t+l-1}) \cdots$$
$$\sum_{\underline{\mathbf{I}}_{l_j} \in \mathcal{I}} \xi_{2l_j-1}(\underline{\mathbf{I}}_{l_j}^r) \cdots \sum_{\underline{\mathbf{I}}_k \in \{\underline{\mathbf{I}}_k^r, \underline{\mathbf{I}}_{k-1}\}} \epsilon_3 \cdots$$
$$\sum_{\underline{\mathbf{I}}_K \in \{\underline{\mathbf{I}}_K^r, \underline{\mathbf{I}}_{K-1}\}} p(\hat{\mathbf{I}}_{t+K}|\mathbf{X}_t, \mathbf{s}_{t+K}, \hat{\mathbf{I}}_{t+K}^r, \hat{\mathbf{I}}_{t+K-1})D_{max}\sqrt{N}K\|\mathbf{X}_t\|$$
$$\sum_{\underline{\mathbf{X}}^K \in \mathcal{R}(\mathbf{X}_t, K, \tilde{A}_{\epsilon_2})} p(\mathbf{X}_{t+1}|\mathbf{Y}_t) \prod_{i=2}^{K} \Big(p(\mathbf{X}_{t+i}|\mathbf{X}_{t+i-1}, \mathbf{s}_{t+i-1}, \mathbf{I}_{t+i-1})\Big)$$
$$< KD_{max}\sqrt{N}(\epsilon_3)^{k_1}(2\epsilon_3)^{k_2}\|\mathbf{X}_t\|, \quad (79)$$

where in the above, summations are nested, and to make the probability expressions shorter, we have shown only r.v.'s and removed their corresponding equality expressions. For example, we have used $\hat{\mathbf{I}}_{t+l}^r$ to represent $\hat{\mathbf{I}}_{t+l}^r = \underline{\mathbf{I}}_l^r$. To obtain the last inequality, we have considered the summations from the innermost to the outermost, and we have used the fact that the summation of probabilities over all possible choices adds up to one, the summation over the error term of the type $\xi_{2i_j-1}$ contributes a multiplicative factor of $\epsilon_3$, the summation over the error of the type $\xi_{2l_j}$ contributes a multiplicative factor of $2\epsilon_3$, and the fact that for any given $\underline{\mathbf{I}}^K$ and $\underline{\mathbf{s}}^K$, the last summation is less than one.

The inequality (79) shows that the obtained upper-bound does not depend on the order of $i_j$'s or $l_j$'s. Since there are $C_K^{k_1}C_K^{k_2}$ number of sub-summations with $k_1$ errors of the type $\xi_{2i-1}$ and with $k_2$ errors of the type $\xi_{2i}$, considering all possible values for $k_1$ and $k_2$, in particular the case where $k_1 = k_2 = 0$ that is accounted for by $\Sigma_m$, and using (78) and (79), we obtain

$$|(73) - \bar{\mathbb{E}}_{\mathbf{X}_t}[g(\mathbf{I}^K, \mathbf{s}^K)|\mathbf{Y}_t]| \leq$$
$$KD_{max}\sqrt{N}\Big(K\epsilon_2 + \big((1+\epsilon_3)^K(1+2\epsilon_3)^K - 1\big)\Big)\|\mathbf{X}_t\|. \quad (80)$$

Combining (72), (73), (75), and (80), we have

$$\Big|\mathbb{E}[g(\mathbf{W}_t^K)|\mathbf{Y}_t] - \bar{\mathbb{E}}_{\mathbf{X}_t}[g(\mathbf{I}_t^K, \mathbf{s}_t^K)|\mathbf{Y}_t]\Big| <$$
$$KD_{max}\sqrt{N}\Big(2K\epsilon_2 + \big((1+2\epsilon_3)^{2K} - 1\big)\Big)\|\mathbf{X}_t\|, \quad (81)$$

which by the definitions of $g(\mathbf{I}_t^K, \mathbf{s}_t^K)$ and $g(\mathbf{W}_t^K)$, given in (70) and (71), respectively, is the same as the inequality in the lemma except that in the left hand side of (81), $K$ should be replaced with $K - 1$. If we consider $K - 1$, in all previous steps, we require $\|\mathbf{X}_t\| \geq B_{\epsilon_2,\epsilon_3}^{(K-1)}$. But $B_{\epsilon_2,\epsilon_3}^K \geq B_{\epsilon_2,\epsilon_3}^{(K-1)}$. On the other hand, the right hand side of (81) is an increasing function of $K$. Therefore, $\|\mathbf{X}_t\| \geq B_{\epsilon_2,\epsilon_3}^K$ is sufficient to have the inequality in the lemma, completing the proof. $\qquad\square$

**Lemma 6.** *Suppose arbitrary positive real numbers $\epsilon_2$, $\epsilon_3$, $\epsilon_4$, and $\epsilon_5$ are given, and $K-1 \geq \max(K_{1,\epsilon_4}, K_{2,\epsilon_5}^{(\gamma)})$, where $K_{1,\epsilon_4}$ and $K_{2,\epsilon_5}^{(\gamma)}$ are defined in Section III-B. If $\|\mathbf{X}_t\| \geq B_{\epsilon_2,\epsilon_3}^K$, where $B_{\epsilon_2,\epsilon_3}^K$ is defined in Lemma 5, the following inequality holds:*

$$
\sum_{k=0}^{K-1} \mathbb{E}\big[\mathbf{X}_{t+k}(\mathbf{D}_{t+k}^* - \mathbf{D}_{t+k})|\mathbf{Y}_t\big]
$$
$$
\leq KC_2 + K^2 C_3 + K\sqrt{N}\epsilon_6\|\mathbf{X}_t\| + K\rho\alpha\delta^{-1}\|\mathbf{X}_t\|
$$
$$
+ K\Upsilon(\mathbf{X}_t)\Big(\zeta' + \frac{(1-\delta)}{\delta}\Psi_{\mathbf{Y}_t}^K\Big),
$$

*where $C_2 = 2\delta^{-1}(1-\delta)C_1$ with $C_1$ defined in Lemma 2, $C_3 = \sqrt{N}D_{max}\|\mathbf{a}\| + 2\delta^{-1}(1-\delta)C_1 + \rho\alpha\delta^{-1}(\|\mathbf{a}\| + \sqrt{N}(\tilde{A}_{max} + D_{max}))$, $\zeta'$ is defined in (22), and*

$$
\epsilon_6 = D_{max}\Big(\delta^{-1}\big(2K\epsilon_2 + \big((1+2\epsilon_3)^{2K} - 1\big)\big)
$$
$$
+ \epsilon_4 + 2\epsilon_5\delta^{-1} + \frac{\delta^{-1}}{K}\Big).
$$

*Proof.* We first focus on each individual term in the summation in the lemma. Let $\mathbb{A}_t$ denote the event that (11) holds at time $t$ for $\mathbf{X} = \mathbf{X}(t)$, $\mathbf{s} = \mathbf{s}(t)$, and $\mathbf{\Gamma} = \mathbf{\Gamma}^r(t)$. Using Property 1 and the fact (14) still holds when $\mathbf{D}'(t-1)$ is replaced by $\mathbf{D}^r(t)$, we obtain

$$
\Delta_{\mathbf{D}_{t+k}} \triangleq \mathbb{E}\big[\mathbf{X}_{t+k}(\mathbf{D}_{t+k}^* - \mathbf{D}_{t+k})|\mathbf{Y}_t\big]
$$
$$
\leq \delta\,\zeta'\,\mathbb{E}\big[\mathbf{X}_{t+k}\mathbf{D}_{t+k}^*|\mathbf{Y}_t, \mathbb{A}_{t+k}\big]
$$
$$
+ \rho\alpha\delta\,\mathbb{E}\big[\|\mathbf{X}_{t+k}\| \mid \mathbf{Y}_t, \mathbb{A}_{t+k}\big]
$$
$$
+ (1-\delta)\mathbb{E}\big[\mathbf{X}_{t+k}(\mathbf{D}_{t+k}^* - \mathbf{D}_{t+k})|\mathbf{Y}_t, \mathbb{A}_{t+k}^c\big]
$$
$$
= \delta\,\zeta'\,\mathbb{E}\big[\mathbf{X}_{t+k}\mathbf{D}_{t+k}^*|\mathbf{Y}_t\big] + \rho\alpha\delta\,\mathbb{E}\big[\|\mathbf{X}_{t+k}\| \mid \mathbf{Y}_t\big]
$$
$$
+ (1-\delta)\mathbb{E}\Big[\mathbf{X}_{t+k}\Big(\mathbf{D}_{t+k}^* - (1-\rho)\mathbf{D}_{t+k-1}'
$$
$$
+ (1-\rho)\mathbf{D}_{t+k-1}' - \mathbf{D}_{t+k}\Big) \,\Big|\, \mathbf{Y}_t, \mathbb{A}_{t+k}^c\Big], \quad (82)
$$

where $\zeta'$ is defined in (22). In the above, we have used the fact that the current and the past queue-lengths, or the past chosen rates, are independent of the event $\mathbb{A}_{t+k}$ (or its complement), and that the event $\mathbb{A}_{t+k}$ occurs with probability $\delta$.

Using (14) and adding and subtracting $\mathbf{D}_{t+k-1}$, we can show that

$$
\Delta_{\mathbf{D}_{t+k}} \leq \delta\,\zeta'\,\mathbb{E}\big[\mathbf{X}_{t+k}\mathbf{D}_{t+k}^*|\mathbf{Y}_t\big]
$$
$$
+ (1-\delta)\mathbb{E}\Big[\mathbf{X}_{t+k}\Big(\mathbf{D}_{t+k}^* - \mathbf{D}_{t+k-1}
$$
$$
+ \mathbf{D}_{t+k-1} - (1-\rho)\mathbf{D}_{t+k-1}'\Big) \,\Big|\, \mathbf{Y}_t, \mathbb{A}_{t+k}^c\Big]
$$
$$
+ \rho\alpha(1-\delta)\mathbb{E}\big[\|\mathbf{X}_{t+k}\| \mid \mathbf{Y}_t, \mathbb{A}_{t+k}^c\big]
$$
$$
+ \rho\alpha\delta\mathbb{E}\big[\|\mathbf{X}_{t+k}\| \mid \mathbf{Y}_t\big].
$$

Similarly, since $\mathbb{A}_{t+k}^c$ is independent of $\mathbf{X}_{t+k}$, by adding and subtracting $\mathbf{D}_{t+k-1}^*$, we obtain

$$
\Delta_{\mathbf{D}_{t+k}} \leq \delta\,\zeta'\,\mathbb{E}\big[\mathbf{X}_{t+k}\mathbf{D}_{t+k}^*|\mathbf{Y}_t\big]
$$
$$
+ (1-\delta)\mathbb{E}\Big[\mathbf{X}_{t+k}\Big(\mathbf{D}_{t+k}^* - \mathbf{D}_{t+k-1}^*
$$
$$
+ \mathbf{D}_{t+k-1}^* - \mathbf{D}_{t+k-1}\Big) \,\Big|\, \mathbf{Y}_t, \mathbb{A}_{t+k}^c\Big]
$$
$$
+ (1-\delta)\mathbb{E}\Big[\mathbf{X}_{t+k}(\mathbf{D}_{t+k-1} - (1-\rho)\mathbf{D}_{t+k-1}')\,\Big|\, \mathbf{Y}_t, \mathbb{A}_{t+k}^c\Big]
$$
$$
+ \rho\alpha\mathbb{E}\big[\|\mathbf{X}_{t+k}\|\,|\,\mathbf{Y}_t\big],
$$

which combined with (5) yields

$$
\Delta_{\mathbf{D}_{t+k}} \leq \delta\,\zeta'\,\mathbb{E}[\mathbf{X}_{t+k}\mathbf{D}_{t+k}^*|\mathbf{Y}_t]
$$
$$
+ (1-\delta)\mathbb{E}\big[\mathbf{X}_{t+k}(\mathbf{D}_{t+k-1} - (1-\rho)\mathbf{D}_{t+k-1}')|\mathbf{Y}_t\big]
$$
$$
+ (1-\delta)\mathbb{E}\big[\mathbf{X}_{t+k}(\mathbf{D}_{t+k}^* - \mathbf{D}_{t+k-1}^*)|\mathbf{Y}_t\big]
$$
$$
+ (1-\delta)\mathbb{E}\Big[\big(\mathbf{A}_{t+k-1} - \mathbf{D}_{t+k-1} + \mathbf{U}_{t+k-1}\big)
$$
$$
(\mathbf{D}_{t+k-1}^* - \mathbf{D}_{t+k-1})|\mathbf{Y}_t\Big]
$$
$$
+ \rho\alpha\mathbb{E}\big[\|\mathbf{X}_{t+k}\|\,|\,\mathbf{Y}_t\big]
$$
$$
+ (1-\delta)\mathbb{E}\big[(\mathbf{X}_{t+k-1})(\mathbf{D}_{t+k-1}^* - \mathbf{D}_{t+k-1})|\mathbf{Y}_t\big]. \quad (83)
$$

By induction and summing over $k$, from (83), we obtain

$$
\sum_{k=0}^{K-1} \mathbb{E}[\mathbf{X}_{t+k}(\mathbf{D}_{t+k}^* - \mathbf{D}_{t+k})|\mathbf{Y}_t]
$$
$$
= \sum_{m=0}^{K-2}\sum_{i=1}^{K-m-1}(1-\delta)^{i-1}\delta\,\zeta'\,\mathbb{E}[\mathbf{X}_{t+m+1}\mathbf{D}_{t+m+1}^*|\mathbf{Y}_t] \quad (84)
$$
$$
+ \sum_{m=0}^{K-2}\sum_{i=1}^{K-m-1}(1-\delta)^i\mathbb{E}\Big[\mathbf{X}_{t+m+1}
$$
$$
(\mathbf{D}_{t+m} - (1-\rho)\mathbf{D}_{t+m}')|\mathbf{Y}_t\Big] \quad (85)
$$
$$
+ \sum_{m=0}^{K-2}\sum_{i=1}^{K-m-1}(1-\delta)^i\mathbb{E}\Big[\mathbf{X}_{t+m+1}
$$
$$
(\mathbf{D}_{t+m+1}^* - \mathbf{D}_{t+m}^*)|\mathbf{Y}_t\Big] \quad (86)
$$
$$
+ \sum_{m=0}^{K-2}\sum_{i=1}^{K-m-1}(1-\delta)^i\mathbb{E}\Big[\big(\mathbf{A}_{t+m} - \mathbf{D}_{t+m} + \mathbf{U}_{t+m}\big)
$$
$$
(\mathbf{D}_{t+m}^* - \mathbf{D}_{t+m})|\mathbf{Y}_t\Big] \quad (87)
$$
$$
+ \rho\alpha\sum_{m=0}^{K-2}\sum_{i=1}^{K-m-1}(1-\delta)^{i-1}\mathbb{E}[\|\mathbf{X}_{t+m+1}\||\mathbf{Y}_t] \quad (88)
$$
$$
+ \frac{(1 - (1-\delta)^K)}{\delta}\Big(\mathbf{X}_t(\mathbf{D}_t^* - \mathbf{D}_t)\Big) \quad (89)
$$

In the sequel, we find an upper bound for each summation term in the above. First, note that similar to (62), we have the following inequality:

$$
\mathbb{E}[\mathbf{X}_{t+m+1}\mathbf{D}_{t+m+1}^*|\mathbf{Y}_t] \leq
$$
$$
\max_{\mathbf{I}\in\mathcal{I}}\mathbb{E}[\mathbf{X}_t\mathbf{D}(\mathbf{s}_{t+m+1}, \mathbf{I})|\mathbf{Y}_t] + \sqrt{N}(m+1)D_{max}\|\mathbf{a}\|. \quad (90)
$$

Using this inequality and that $\delta\sum_{i=1}^{K-m-1}(1-\delta)^i < 1$, we

have

$$
\begin{aligned}
(84) &\leq \zeta' \sum_{m=0}^{k-2} \sum_{\mathbf{s}\in\mathcal{S}} \mathbb{E}[\mathbf{X}_t \mathbf{D}^*(\mathbf{X}_t,\mathbf{s})\mathbf{1}_{\mathbf{s}_{t+m+1}=\mathbf{s}}|\mathbf{Y}_t] \\
&\quad - (K-1)\zeta' \sum_{\mathbf{s}\in\mathcal{S}} \pi(\mathbf{s})\mathbf{X}_t \mathbf{D}^*(\mathbf{X}_t,\mathbf{s}) \\
&\quad + (K-1)\zeta' \sum_{\mathbf{s}\in\mathcal{S}} \pi(\mathbf{s})\mathbf{X}_t \mathbf{D}^*(\mathbf{X}_t,\mathbf{s}) + K^2\sqrt{N}D_{max}\|\mathbf{a}\| \\
&\leq \zeta'(K-1)\sum_{\mathbf{s}\in\mathcal{S}} \\
&\quad \Big( \mathbb{E}\big[\frac{1}{K-1}\sum_{m=0}^{K-2}\mathbf{1}_{\mathbf{s}_{t+m+1}=\mathbf{s}}|\mathbf{Y}_t\big] - \pi(\mathbf{s})\Big)\mathbf{X}_t\mathbf{D}^*(\mathbf{X}_t,\mathbf{s}) \\
&\quad + (K-1)\zeta' \mathbb{E}[\mathbf{X}_t\mathbf{D}^*(\mathbf{X}_t,\mathbf{s})] + K^2\sqrt{N}D_{max}\|\mathbf{a}\| \\
&< K\epsilon_4\|\mathbf{X}_t\|\sqrt{N}D_{max} + K\zeta'\Upsilon(\mathbf{X}_t) + K^2\sqrt{N}D_{max}\|\mathbf{a}\|,
\end{aligned}
$$
(91)

where the last inequality follows from the assumption that $K \geq K_{1,\epsilon_4}$, the inequality $\zeta' \leq 1$, the definition in (16), and that $D_{max}$ is the transmission rate upper-bound.

To derive an upper-bound for (85), we can use (5) and the inequality $\sum_{i=1}^{K-m-1}(1-\delta)^i < \delta^{-1}(1-\delta)$ to show that

$$
\begin{aligned}
(85) &\leq \delta^{-1}(1-\delta)\sum_{m=0}^{K-2}\mathbb{E}\Big[\big(\sum_{l=0}^{m}\mathbf{A}_l - \sum_{l=0}^{m}(\mathbf{D}_l - \mathbf{U}_l)\big) \\
&\qquad\qquad (\mathbf{D}_{t+m} - (1-\rho)\mathbf{D}'_{t+m})|\mathbf{Y}_t\Big] \\
&\quad + \delta^{-1}(1-\delta)\sum_{m=0}^{K-2}\mathbb{E}\big[\mathbf{X}_t(\mathbf{D}_{t+m} - (1-\rho)\mathbf{D}'_{t+m})|\mathbf{Y}_t\big] \\
&\leq \delta^{-1}(1-\delta)\sum_{m=0}^{K-2}(m+1)\big(\sqrt{N}D_{max}\|\mathbf{a}\| + (1-\rho)D_{max}^2 N\big) \\
&\quad + \delta^{-1}(1-\delta)\sum_{m=0}^{K-2}\mathbb{E}\big[\mathbf{X}_t(\mathbf{D}_{t+m} - (1-\rho)\mathbf{D}'_{t+m})|\mathbf{Y}_t\big].
\end{aligned}
$$
(92)

Since by assumption $\|\mathbf{X}_t\| \geq B_{\epsilon_2,\epsilon_3}^K$, we can use Lemma 5 to replace the expectation in the above with an expectation of the type $\bar{\mathbb{E}}_{\mathbf{X}_t}[\cdot]$, leading to

$$
\begin{aligned}
(85) &\leq K^2\delta^{-1}(1-\delta)C_1 + \\
&\quad + \delta^{-1}(1-\delta)\sum_{m=0}^{K-2}\bar{\mathbb{E}}_{\mathbf{X}_t}[\mathbf{X}_t(\mathbf{D}_{t+m} - (1-\rho)\mathbf{D}'_{t+m})|\mathbf{Y}_t] \\
&\quad + K\delta^{-1}(1-\delta)D_{max}\sqrt{N} \\
&\qquad\qquad \Big(2K\epsilon_2 + \big((1+2\epsilon_3)^{2K}-1\big)\Big)\|\mathbf{X}_t\|.
\end{aligned}
$$
(93)

We now focus on (86). Let $\mathbf{D}_t^{*f} \triangleq \mathbf{D}^*(\mathbf{X}_{t+1},\mathbf{s}_t)$; hence, $\mathbf{D}_t^{*f}$ is the optimal rate corresponding to the queue-length vector at time-slot $t+1$ but the channel state at time $t$. Using this definition and Lemma 2, we can show that

$$
\begin{aligned}
(86) &= \sum_{m=0}^{K-2}\gamma_{K-m-1}\mathbb{E}\Big[\mathbf{X}_{t+m+1}\big(\mathbf{D}_{t+m+1}^* \\
&\qquad\qquad - \mathbf{D}_{t+m}^{*f} + \mathbf{D}_{t+m}^{*f} - \mathbf{D}_{t+m}^*\big)|\mathbf{Y}_t\Big] \\
&\leq \sum_{m=0}^{K-2}\gamma_{K-m-1}\Big(C_1 + \underbrace{\mathbb{E}\big[\mathbf{X}_{t+m+1}(\mathbf{D}_{t+m+1}^* - \mathbf{D}_{t+m}^{*f})|\mathbf{Y}_t\big]}_{\Delta^{*f}}\Big),
\end{aligned}
$$
(94)

where $\gamma_{K-m-1}$ belongs to the sequence $\gamma$ that is defined in Section III-B. We first find an upper-bound for $\Delta^{*f}$. Let $\mathbf{Y}_t^{ext} \triangleq (\mathbf{Y}_t,\mathbf{s}_{t+m},\mathbf{s}_{t+m+1})$. Using Lemma 3, we have

$$
\begin{aligned}
\Delta^{*f} &= \sum_{(\mathbf{s}_1,\mathbf{s}_2)\in\mathcal{S}^2}\sum_{\mathbf{X}}\underbrace{p\Big((\mathbf{s}_{t+m},\mathbf{s}_{t+m+1})=(\mathbf{s}_1,\mathbf{s}_2)|\mathbf{Y}_t\Big)}_{\tilde{p}(\mathbf{s}_1,\mathbf{s}_2)} \\
&\qquad p\Big(\mathbf{X}_{t+m+1}=\mathbf{X}|\mathbf{Y}_t^{ext}\Big)\,\mathbf{X}\big(\mathbf{D}^*(\mathbf{X},\mathbf{s}_2)-\mathbf{D}^*(\mathbf{X},\mathbf{s}_1)\big)\Big] \\
&\leq \sum_{(\mathbf{s}_1,\mathbf{s}_2)\in\mathcal{S}^2}\sum_{\mathbf{X}}\Big[\tilde{p}(\mathbf{s}_1,\mathbf{s}_2)p(\mathbf{X}_{t+m+1}=\mathbf{X}|\mathbf{Y}_t^{ext}) \\
&\qquad \big(\mathbf{X}_t(\mathbf{D}^*(\mathbf{X}_t,\mathbf{s}_2)-\mathbf{D}^*(\mathbf{X}_t,\mathbf{s}_1)) \\
&\qquad + (\mathbf{X}-\mathbf{X}_t)(\mathbf{D}^*(\mathbf{X},\mathbf{s}_2)-\mathbf{D}^*(\mathbf{X}_t,\mathbf{s}_1))\big)\Big].
\end{aligned}
$$

Replacing the distribution over the queue-length vector with the distribution over arrivals and transmission rates, which leads to the same expectation, and letting

$$
\begin{aligned}
&p(\{\underline{\mathbf{A}}_i\}_{i=0}^m,\{\underline{\mathbf{D}}_i\}_{i=0}^m|\mathbf{Y}_t^{ext}) = \\
&p(\{\mathbf{A}_{t+i}\}_{i=0}^m=\{\underline{\mathbf{A}}_i\}_{i=0}^m,\{\mathbf{D}_{t+i}\}_{i=0}^m=\{\underline{\mathbf{D}}_i\}_{i=0}^m|\mathbf{Y}_t^{ext}),
\end{aligned}
$$

we can rewrite the above inequality as

$$
\begin{aligned}
\Delta^{*f} &= \Big[\sum_{(\mathbf{s}_1,\mathbf{s}_2)\in\mathcal{S}^2}\tilde{p}(\mathbf{s}_1,\mathbf{s}_2)\mathbf{X}_t\big(\mathbf{D}^*(\mathbf{X}_t,\mathbf{s}_2)-\mathbf{D}^*(\mathbf{X}_t,\mathbf{s}_1)\big) \\
&\qquad\qquad \sum_{\mathbf{X}}p\big(\mathbf{X}_{t+m+1}=\mathbf{X}|\mathbf{Y}_t^{ext}\big)\Big] \\
&\quad + \Big[\sum_{(\mathbf{s}_1,\mathbf{s}_2)\in\mathcal{S}^2}\sum_{\{\underline{\mathbf{A}}_i\}_{i=0}^m,\{\underline{\mathbf{D}}_i\}_{i=0}^m}\tilde{p}(\mathbf{s}_1,\mathbf{s}_2) \\
&\qquad\qquad p\big(\{\underline{\mathbf{A}}_i\}_{i=0}^m,\{\underline{\mathbf{D}}_i\}_{i=0}^m|\mathbf{Y}_t^{ext}\big) \\
&\quad \big(\sum_{i=0}^m\underline{\mathbf{A}}_i - \sum_{i=0}^m(\underline{\mathbf{D}}_i - \underline{\mathbf{U}}_i)\big)\big(\mathbf{D}^*(\underline{\mathbf{X}}_{t+m+1},\mathbf{s}_2)-\mathbf{D}^*(\mathbf{X}_t,\mathbf{s}_1)\big)\Big] \\
&\leq \sum_{\mathbf{s}\in\mathcal{S}}p(\mathbf{s}_{t+m+1}=\mathbf{s}|\mathbf{Y}_t)\mathbf{X}_t\mathbf{D}^*(\mathbf{X}_t,\mathbf{s}) \\
&\quad - \sum_{\mathbf{s}\in\mathcal{S}}p(\mathbf{s}_{t+m}=\mathbf{s}|\mathbf{Y}_t)\mathbf{X}_t\mathbf{D}^*(\mathbf{X}_t,\mathbf{s}) + (m+1)C_1,
\end{aligned}
$$
(95)

where the term $(m+1)C_1$, with $C_1$ defined in Lemma 2, is an upper-bound for the second bracketed summation in the above. Note that in the above, $\underline{\mathbf{X}}_{t+m+1}$ represents one realization of the vector $\mathbf{X}_{t+m+1}$ and is determined by $\mathbf{X}_t$ and the sequences $\{\underline{\mathbf{A}}_i\}_{i=0}^m$ and $\{\underline{\mathbf{D}}_i\}_{i=0}^m$.

Now, let $S_\gamma^K = \sum_{m=0}^{K-2}\gamma_{K-m-1}$, for which we have $S_\gamma^K < K\delta^{-1}(1-\delta)$. Using this definition and the assumption that $K-1 \geq K_{2,\epsilon_5}^{(\gamma)}$, we obtain

$$
\begin{aligned}
&\Big| \sum_{m=0}^{K-2}\gamma_{K-m-1}\sum_{\mathbf{s}\in\mathcal{S}}p(\mathbf{s}_{t+m+1}=\mathbf{s}|\mathbf{Y}_t)\mathbf{X}_t\mathbf{D}^*(\mathbf{X}_t,\mathbf{s}) \\
&\quad - \sum_{m=0}^{K-2}\gamma_{K-m-1}\sum_{\mathbf{s}\in\mathcal{S}}\pi(\mathbf{s})\mathbf{X}_t\mathbf{D}^*(\mathbf{X}_t,\mathbf{s})\Big| \\
&= (S_\gamma^K)\Big|\sum_{\mathbf{s}\in\mathcal{S}}\Big(\mathbb{E}\big[\frac{1}{(S_\gamma^K)}\sum_{m=0}^{K-2}\gamma_{K-m-1}\mathbf{1}_{\mathbf{s}_{t+m+1}=\mathbf{s}}|\mathbf{Y}_t\big]-\pi(\mathbf{s})\Big) \\
&\qquad\qquad\qquad\qquad\qquad \mathbf{X}_t\mathbf{D}^*(\mathbf{X}_t,\mathbf{s})\Big| \\
&< K\epsilon_5\delta^{-1}(1-\delta)\sqrt{N}D_{max}\|\mathbf{X}_t\|.
\end{aligned}
$$
(96)

We can obtain a similar inequality as the one in (96) with $\mathbf{s}_{t+m+1}$ is replaced with $\mathbf{s}_{t+m}$, which along with the results from (94), (95), and (96), yields

$$(86) \le (K + K^2)\delta^{-1}(1-\delta)C_1$$
$$+ 2K\epsilon_5\delta^{-1}(1-\delta)\sqrt{N}D_{max}\|\mathbf{X}_t\|. \quad (97)$$

Remaining terms that should be upper-bounded are (87), (88), and (89). Using the triangle inequality and the fact that $(\mathbb{E}[\|\mathbf{X}_{t+m+1} - \mathbf{X}_t\| \ |\mathbf{Y}_t])^2 \le \mathbb{E}[\|\mathbf{X}_{t+m+1} - \mathbf{X}_t\|^2|\mathbf{Y}_t]$, we can show that

$$\mathbb{E}[\|\mathbf{X}_{t+m+1}\| \ |\mathbf{Y}_t] \le \|\mathbf{X}_t\| + \sqrt{\mathbb{E}[\|\mathbf{X}_{t+m+1} - \mathbf{X}_t\|^2 \ |\mathbf{Y}_t]}$$
$$\le \|\mathbf{X}_t\| + \sqrt{m+1}\sqrt{\mathbb{E}[\|\mathbf{A}\|^2]} + (m+1)\|\mathbf{a}\|$$
$$+ \sqrt{N}(m+1)D_{max}.$$

Hence,

$$(88) \le \rho\alpha\delta^{-1}K\|\mathbf{X}_t\|$$
$$+ \rho\alpha\delta^{-1}K^2\big(\|\mathbf{a}\| + \sqrt{N}(\tilde{A}_{max} + D_{max})\big), \quad (98)$$

where $\tilde{A}_{max}$ is defined in (3). Using the same methods, we can easily show that

$$(87) \le K\delta^{-1}(1-\delta)C_1, \quad (89) \le \delta^{-1}\sqrt{N}D_{max}\|\mathbf{X}_t\|. \quad (99)$$

Combining the results from (84)-(89), (91), (93), (97), (98), and (99), we obtain the inequality in the lemma, as required. ∎

**Lemma 7.** *Suppose arbitrary positive real numbers $\epsilon_2'$, $\epsilon_3'$, and $\epsilon_4'$ are given. Suppose $K \ge K_{1,\epsilon_4'}$, where $K_{1,\epsilon_4'}$ is defined in Section III-B. Then, if $\|\mathbf{X}_t\| \ge B^K_{\epsilon_2',\epsilon_3'}$, where $B^K_{\epsilon_2',\epsilon_3'}$ is defined in Lemma 5, the following inequality holds:*

$$\sum_{k=0}^{K-1} \mathbb{E}[\mathbf{X}_{t+k}(\mathbf{D}^*_{t+k} - \mathbf{D}_{t+k})|\mathbf{Y}_t]$$
$$\le K^2 C_3' + K\sqrt{N}\epsilon_6'\|\mathbf{X}_t\| + K\Upsilon(\mathbf{X}_t)(1 - \Phi^K_{\mathbf{Y}_t}),$$

*where $C_3' = C_1$ with $C_1$ defined in Lemma 2 and*

$$\epsilon_6' = D_{max}\Big(2K\epsilon_2' + \big((1 + 2\epsilon_3')^{2K} - 1\big) + \epsilon_4'\Big).$$

*Proof.* The proof is a similar but a simpler version of the proof given for Lemma 5. ∎